\documentclass[twocolumn,showpacs,showkeys,amsmath,amssymb]{revtex4}

\usepackage{graphicx}

\begin{document}

\title{Optical caustics of Kerr spacetime: the full structure}

\author{V. Bozza$^{a,b}$}

\affiliation{$^a$ Dipartimento di Fisica ``E.R. Caianiello'',
Universit\`a di Salerno, via Allende,
I-84081 Baronissi (SA), Italy.\\
  $^b$  Istituto Nazionale di Fisica Nucleare, Sezione di
 Napoli. }

\date{\today}

\begin{abstract}
We present an exhaustive numerical investigation of the optical
caustics in gravitational lensing by a spinning black hole for an
observer at infinity. Besides the primary caustic, we examine
higher order caustics, formed by photons performing one or several
loops around the black hole. Our investigation covers the whole
parameter space, including the black hole spin, its inclination
with respect to the line of sight, the source distance, and the
caustic order. By comparing our results with the available
analytical approximations, we find perfect agreement in their
respective domains of validity. We then prove that all caustics
maintain their shape (a tube with astroidal cross-section) in the
entire parameter space without suffering any transitions to
different caustic shapes. For nearly extremal spin, however,
higher order caustics grow so large that their cross-sections at
fixed radii wind several times around the black hole. As a
consequence, for each caustic order, the number of images ranges
from 2 to $2(n+1)$, where $n$ is the number of loops spanned by
the caustic. As for the critical curves, we note that for high
values of the spin they develop a small dip on the side
corresponding to prograde orbits.
\end{abstract}



\maketitle

\section{Introduction}

If General Relativity is the correct theory of gravity, the Kerr
solution describes the spacetime metric outside spinning black
holes \cite{Kerr}. Therefore it is currently utilized in all
models trying to reconstruct the phenomena observed around
observed astrophysical black holes, either remnants of stellar
collapse or supermassive black holes lying in the central regions
of several galaxies \cite{Cha,Mel}. A crucial step in the
comprehension of physics in such extreme environments is the
complete understanding of the phenomenology related to the bending
of photon trajectories caused by spacetime curvature. In order to
get reliable predictions on any observables, it is necessary to
keep in mind that a Kerr black hole acts as a very strong
gravitational lens, generating an infinite number of images of the
same source \cite{HasPer}. The total flux of a source close to a
Kerr black hole, (such as the accretion disk itself
\cite{Accret,BecDon}, an isolated bright spot on it
\cite{KVP,BroLoe} or a star orbiting the black hole \cite{CunBar})
gets a significant contribution from the secondary image and from
higher order images \cite{BecDon}. This is also true for the
details of fine structures in the profile of spectral lines, such
as the iron $K\alpha$ line in the X-ray domain, which are strong
indicators of the presence of an intrinsic angular momentum of the
black hole \cite{Tan}.

On the other hand, the progresses in radio \cite{Kri} and infrared
band \cite{IR} interferometry, along with the projects of X-ray
interferometry in space (MAXIM, http://maxim.gsfc.nasa.gov),
foreshadow that resolved pictures of the nearest supermassive
black hole (Sgr A* in the center of the Milky Way) will be
feasible in a not so far future \cite{FMA}. This will represent a
spectacular advance in our knowledge of black hole physics. In
particular, the contributions of different images of the same
source will be identified and studied separately. Higher order
images will then provide a huge amount of independent information
on the inner portions of the accretion disks of supermassive black
holes and will become precious witnesses of the strong
gravitational field just outside the horizon.

Gravitational lensing theory states that the multiplicity of the
images of a given source depends on the source-lens-observer
configuration. In a given spacetime metric, for an observer in a
particular spacetime point, the multiplicity only depends on the
source position. If the metric is stationary (as in the case of
the Kerr metric) and the observer is static, the caustic can be
defined in the 3-dimensional subspace at constant time as a
2-dimensional surface separating regions of space in which a
source would give rise to a different number of images. When a
point-like source crosses a caustic, a pair of additional images
with infinite magnification is created or destroyed (the finite
source size acts as a cut-off for real sources) \cite{SEF}.

It can be easily guessed that the study of the shape of the
caustics is of fundamental importance for a reliable and complete
description of the whole phenomenology related to the environment
of astrophysical black holes. In fact, the multiplicity of the
images and their brightness is essentially determined by the
position of the source within the caustic structure. Even temporal
variations in the observed overall luminosity may be due to
caustic crossing of bright features around the black hole
\cite{RauBla}.

Surprisingly, 45 years after the discovery of the Kerr metric, the
complete structure of the caustics of a spinning black hole has
not yet been derived. Indeed, the complexity of the metric
prevents from finding simple analytical solutions for the caustic
surfaces. Even numerical studies are very challenging and not
straightforward. The first indication of the existence of
non-degenerate caustics came from the work of Cunningham and
Bardeen \cite{CunBar}, who traced the light curves of a source
star orbiting a spinning black hole. They noticed that the
magnification of the primary and secondary images diverged at some
particular points, signaling that caustic crossings were
occurring. In spite of the huge number of ray-tracing codes in
Kerr spacetime developed in so many years, the only comprehensive
study of the caustic surfaces has been performed by Rauch and
Blandford \cite{RauBla}. They have explicitly shown that the
primary caustic is a tube with a cross-section having the shape of
an astroid (a closed curve with four cusps), which is very typical
in gravitational lensing theory as soon as the spherical symmetry
of the lens is broken by an external or internal perturbation.
They have shown several pictures of the primary caustic and
derived some simple asymptotic behaviors for its size. Besides the
primary caustic, the authors have mentioned the existence of
higher order caustics, but they have not shown any pictures of
them, leaving several questions about the size and the shape of
these caustics open. Later on, Sereno and De Luca have found an
analytical approximation for the primary caustic valid for large
source distances \cite{SerDeL}. In a series of papers based on the
strong deflection limit approximation \cite{Boz1}, we have derived
a perturbative analytical approximation describing the higher
order caustics (but not the primary caustic)
\cite{BDSS,BDS,BozSca}. These approximations show that at low spin
values the higher order caustics are still tubes with astroidal
cross sections with increasing size. Basically, this is all we
know about caustics in the Kerr spacetime.

All these studies provide several hints about the caustic
structure of the Kerr black hole lens in several limits. Yet, the
fate of the higher order caustics at high values of the spin still
remains unclear. As they become larger and larger with increasing
spin, do they undergo any transitions to different caustic shapes?
Does their size stay finite? Do they merge? The present work
provides clear answers to these and other questions of theoretical
and observational relevance, clarifying the whole panorama of the
caustic structure of the Kerr spacetime. We present a thorough
numerical analysis of the caustics generated by a Kerr black hole
at all caustic orders, studying their dependence on the black hole
spin, its inclination and the source distance. The reliability of
our results is also double-checked against former studies and all
analytical approximations available up to now.

The paper is organized as follows. Section II traces the
methodology followed for the generation of the caustics, referring
to the appendix for a detailed explanation of all steps. Section
III deals with the primary caustic. Section IV discusses the
dependence of higher order caustics on the spin and its
inclination. Section V focuses on the caustics of extremal black
holes. Section VI is devoted to critical curves in the observer's
sky. Section VII contains the conclusions.

\section{Methodology}

The Kerr metric in Boyer-Lindquist coordinates \cite{BoyLin} is
($G=c=1$)
\begin{eqnarray}
& ds^2=&\left( 1 -\frac{2M r}{\rho^2} \right)d
t^2-\frac{\rho^2}{\Delta} dr^2-\rho^2 d\vartheta^2 \nonumber \\
&& - \left[ \Delta + \frac{2M r(r^2+a^2)}{\rho^2} \right] \sin^2
\vartheta d\phi^2 \nonumber
\\&&+\frac{4aMr
\sin^2\vartheta}{\rho^2} dt d\phi \\%
& \Delta=&r^2-2Mr+a^2, \\%
& \rho^2=& r^2+a^2 \cos^2\vartheta.
\end{eqnarray}

$M$ is the mass of the black hole and $a$ is the specific angular
momentum, ranging from 0 (Schwarzschild black hole) to $M$
(extremal black hole). The horizon radius is $r_h =
M+\sqrt{M^2-a^2}$.

Let us consider a static observer at coordinates
$(r_o,\vartheta_o, \phi_o)$ with $\phi_o=0$ and let us assume $r_o
\gg 2M$ for simplicity. In practice, all astrophysical situations
satisfy this limit. The observer's polar coordinate $\vartheta_o$
also coincides with the inclination of the spin axis with respect
to the line of sight between the observer and the black hole. It
is often convenient to work with the coordinate $\mu \equiv \cos
\vartheta$ instead of $\vartheta$. Therefore, one can define
$\mu_o \equiv \cos \vartheta_o$.

Similarly, we consider a source at coordinates $(r_s,\vartheta_s,
\phi_s)$, with $\mu_s\equiv \cos \vartheta_s$.

The observer can construct his own coordinates $(\theta_1,
\theta_2)$ in the sky, such that the black hole is in the origin
and the $\theta_2$ axis coincides with the projection of the spin
of the black hole on the sky.

Now we make a qualitative summary of the numerical algorithm for
finding the caustics and state a few basic equations that should
be sufficient to understand the main points. In Appendix A we
report all the details of the calculation and the definitions of
all variables.

\begin{figure}
\resizebox{\hsize}{!}{\includegraphics{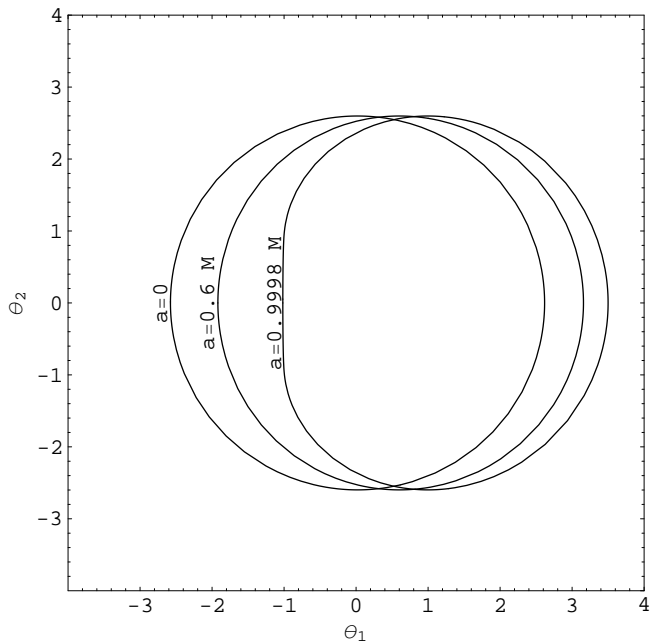}} \caption{The
shadow border as seen by an equatorial observer for different
values of the black hole spin. $\theta_1$ and $\theta_2$ are the
angular coordinates in the observer's sky in units of $2M/r_o$.
The black hole is in the origin and the spin axis points in the
positive $\theta_2$ direction. }
 \label{Fig shadow}
\end{figure}

The first important concept to be reviewed is the shadow of the
black hole \cite{Cha}. Imagine tracing a photon trajectory from
the observer back toward the black hole. The trajectory is
completely determined by the angles $\theta_1$ and $\theta_2$,
which specify the final direction of the photon when it hits the
observer. These angles may be considered as the initial conditions
for tracing back the photon geodesic in the Kerr metric. Then, it
is well-known that there exists a set of photon trajectories that
are traced back to grazing unstable fixed radii orbits around the
black hole. These particular trajectories define a locus in
$(\theta_1,\theta_2)$, which is called the shadow border of the
black hole. As evident from Fig. \ref{Fig shadow}, in the limit
$a\rightarrow 0$, this locus has a circular shape with angular
radius $3\sqrt{3}M/r_o$. For non-vanishing spin, it is slightly
displaced toward the right direction (remember that we have chosen
coordinates such that the spin of the black hole is projected on
the positive $\theta_2$ axis) and becomes slightly flattened on
the left side. The shape of the shadow depends on the black hole
spin and its inclination on the line of sight and can be used, in
principle, to extract these parameters \cite{Cha,FMA,Zak}. The
reason for the name ``shadow border'' is that all sources lying
outside the unstable circular orbits generate images appearing
outside this locus in the observer's sky. Only sources very close
to the black hole may have images inside the shadow border.
Therefore, to an instrument with the sufficient angular
resolution, the black hole should appear as a dark spot (shadow)
surrounded by a more luminous region \cite{FMA}. In the appendix,
we review the derivation of the shape of the shadow border, which
is explicitly given by Eqs. (\ref{Shadow1}) and (\ref{Shadow2}) in
parametric form as functions of $r_m$, which represents the radius
of the unstable photon orbit from which the photon emerges and
hits the observer at angles $(\theta_1,\theta_2)$.

It is convenient to switch from $r_m$ to a more friendly parameter
that we call $\eta$ as defined by Eq. (\ref{rmeta}). $\eta$ is
very similar to an angular variable: as it ranges in $[-\pi,\pi]$,
the whole shadow border is spanned. The intersections with the
$\theta_1$ axis are at $\eta=0$ (with $\theta_1>0$) and $\eta=\pm
\pi$ (with $\theta_1<0$). $\eta=\pi/2$ corresponds to a point
close to the positive $\theta_2$ axis and $\eta=-\pi/2$
corresponds to a point close to the negative $\theta_2$ axis. As
the spin axis is projected to the positive $\theta_2$ axis, we can
deduce that photons reaching the observer from the left of the
black hole ($\eta\simeq\pm \pi$) come from orbits co-rotating with
the black hole, whereas photons reaching the observer from the
right of the black hole ($\eta\simeq0$) come from orbits rotating
in the opposite sense with respect to the black hole. We shall
often refer to photons on co-rotating orbits as prograde photons
and to photons on counter-rotating orbits as retrograde photons.
Photons reaching the observer from above or below the black hole
($\eta\simeq \pm \pi/2$) come from quasi-polar orbits.

The shadow border can be used to construct new coordinates
$(\psi,\eta)$ in the observer's sky, such that the sky coordinates
$(\theta_1,\theta_2)$ can be re-expressed in terms of them
\begin{eqnarray}
&& \theta_1=\theta_1(\psi,\eta) \label{Simth1par}\\
&& \theta_2=\theta_2(\psi,\eta). \label{Simth2par}
\end{eqnarray}
These coordinates are similar to polar coordinates, with $\eta$
playing the role of the polar angle and $\psi$ related to the
modulus $\sqrt{\theta_1^2+\theta_2^2}$. The choice of the precise
form of the radial variable $\psi$ as a function of
$\sqrt{\theta_1^2+\theta_2^2}$ is done with the purpose of
simplifying the numerical calculations, as will be explained in
the following paragraphs.

A detailed study of the geodesics equations at fixed source
distance $r_s$ (see Appendix) provides the sought after relation
between the source's angular coordinates and the variables $\psi$
and $\eta$
\begin{eqnarray}
&& \mu_s=\mu_s(\psi,\eta)  \label{SimLens1}\\
&& \phi_s=\phi_s(\psi,\eta). \label{SimLens2}
\end{eqnarray}
As $\psi$ and $\eta$ determine the position of the images in the
observer's sky, Eqs. (\ref{SimLens1}) and (\ref{SimLens2})
represent the Kerr lens mapping.

Of course, the detailed expression of this mapping may become very
involved and cumbersome to numerical calculations if one does not
make a proper choice of the variable $\psi$. Luckily, the explicit
form of the function $\mu_s$ in Eq. (\ref{SimLens1}) consists of
an oscillating Jacobi elliptic function whose argument
monotonically increases when $\sqrt{\theta_1^2+\theta_2^2}$
decreases. From the physical point of view, this means that the
closer the photon passes to the black hole the larger is the
number of oscillations in the polar motion it performs. Then, it
is natural to define $\psi$ as the argument of this Jacobi
elliptic function. With this definition, the function
$\mu_s(\psi,\eta)$ resembles a sinusoid in $\psi$, with the
amplitude of the oscillations determined by $\eta$.

With a suitable normalization of $\psi$, we can set the period of
the Jacobi elliptic function to $\Delta\psi=2$. In this way, we
can establish the equivalence $\mu_s=0 \Leftrightarrow \psi = m$,
with $m$ integer. Furthermore, the number of inversions in the
polar motion of the photon from the source to the observer is
$m=[\psi+1/2]$, where by $[x]$ we indicate the integer part of
$x$. The precise definition of $\psi$ is given by Eq.
(\ref{psidef}), though it is necessary to go through a conspicuous
portion of the appendix to understand all the details.

As long as the Jacobian determinant of the lens mapping is
different from zero, the mapping is locally invertible and the
number of images stays constant. Images can be created or
destroyed only in the critical points where the Jacobian
determinant vanishes. It turns out that for each value of $m$
there is one Jacobian critical point in the interval between two
consecutive polar inversions $m-1/2<\psi<m+1/2$, for all values of
$\eta$. Therefore, we can identify $m$ with the caustic order. The
primary caustic is obtained with a single polar inversion ($m=1$),
the second order caustic is generated by photon trajectories with
two polar inversions ($m=2$), and so on. With the critical points
found in the $(\psi,\eta)$ space, we can readily draw the critical
curves in the observer's sky by applying Eqs. (\ref{Simth1par})
and (\ref{Simth2par}), and the caustics by applying Eqs.
(\ref{SimLens1}) and (\ref{SimLens2}).

In conclusion, once we fix all the parameters (black hole spin
$a$, spin inclination $\mu_o$, source distance $r_s$, caustic
order $m$), we are able to trace the corresponding caustic and
critical curve through suitable numerical calculations.

\section{Primary caustic}

The primary caustic is the caustic generated by light rays with a
single inversion point in the polar motion. In the Schwarzschild
limit ($a\rightarrow 0$), it degenerates to a line starting from
the black hole and extending to infinity in the opposite direction
with respect to the observer ($\phi=- \pi$). Distant sources that
approach the primary caustic experience the standard weak
deflection gravitational lensing as described in classical
textbooks. The two weak deflection images merge when the source
lies right on the primary caustic, forming the well-known Einstein
ring.

\begin{figure}
\resizebox{\hsize}{!}{\includegraphics{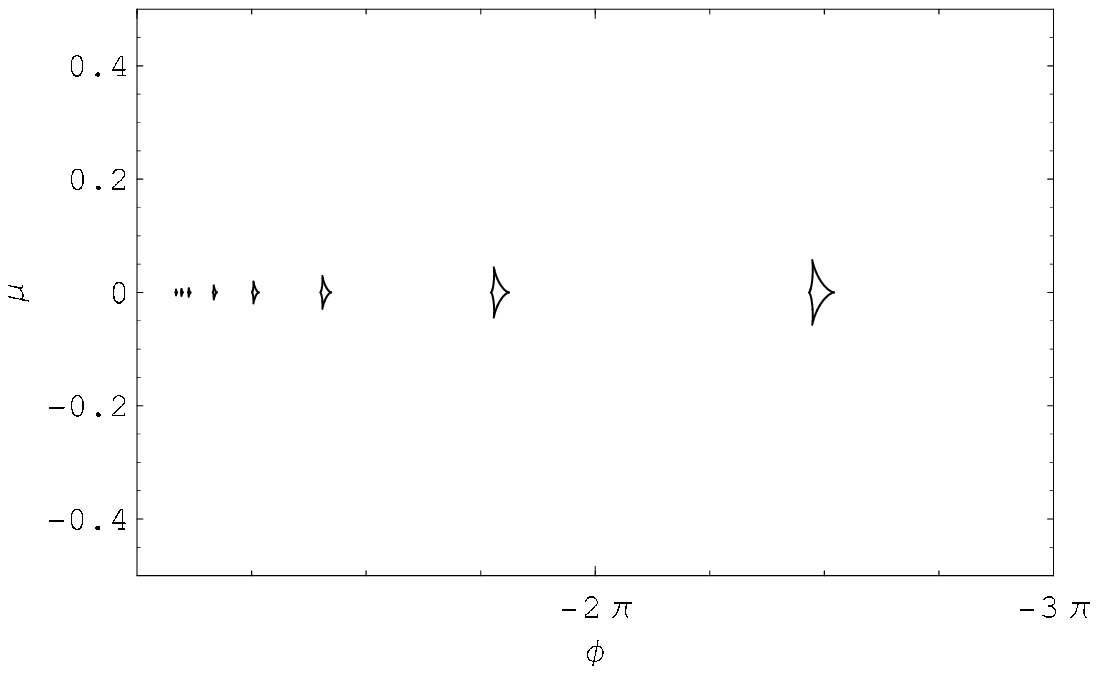}}
\caption{Cross sections of the primary caustic for $a=0.9998M$ and
$\mu_o=0$. From left to right, the radial coordinate is $r_s=5M,
4.5M, 4M, 3M, 2.3M, 1.82M, 1.42M, 1.22M$.}
 \label{Fig WFcross}
\end{figure}

If we turn the black hole spin on, the degenerate one-dimensional
caustic becomes a finite-thickness tube with a cross-section
having the shape of a four-cusped astroid. Several cross-sections
at fixed radii are shown in Fig. \ref{Fig WFcross} for a nearly
extremal Kerr black hole. If the source lies inside the enclosed
region, two additional images are present

\begin{figure}
\resizebox{5.86cm}{!}{\includegraphics{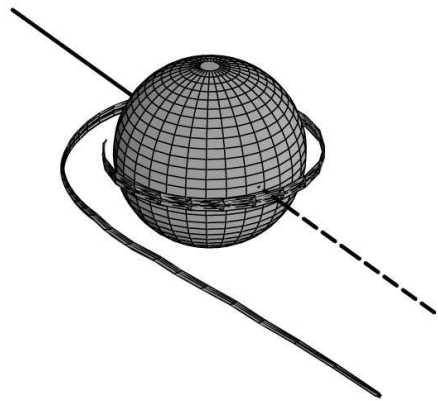}} \caption{The
primary caustic tube for $a=0.9998M$ and $\mu_o=0$. In building
this 3-dimensional representation, we have chosen coordinates such
that $x=r \sin \vartheta \cos \phi$, $y=r \sin \vartheta \sin
\phi$, $z=r \cos \vartheta$, following Ref. \cite{RauBla}. The
spin axis is directed toward the top. The straight solid line
indicates the direction towards the observer, whereas the dashed
line points in the opposite direction. The primary caustic for a
Schwarzschild black hole ($a=0$) coincides with this dashed line.}
 \label{Fig WF3D}
\end{figure}

As already shown in the extensive study by Rauch \& Blandford
\cite{RauBla}, the cross-section is reflection-symmetric at very
large distances, but becomes more and more distorted approaching
the black hole. Moreover, the caustic tube is shifted clockwise
(i.e. in the sense opposite to the rotation of the black hole), as
evident from Fig. \ref{Fig WFcross}, where it can be noted that
the caustic does not lie at $\phi=-\pi$, as in the Schwarzschild
case (We recall that we have put the observer in $\phi_o=0$). This
shift becomes larger and larger for sources closer to the horizon.
A full 3-dimensional picture of the primary caustic tube was first
presented in Ref. \cite{RauBla} and is shown again here in Fig.
\ref{Fig WF3D}. It can be noted that the caustic tube becomes
belt-like as it winds around the black hole horizon, represented
by the spherical surface in the figure.

\begin{figure}
\resizebox{\hsize}{!}{\includegraphics{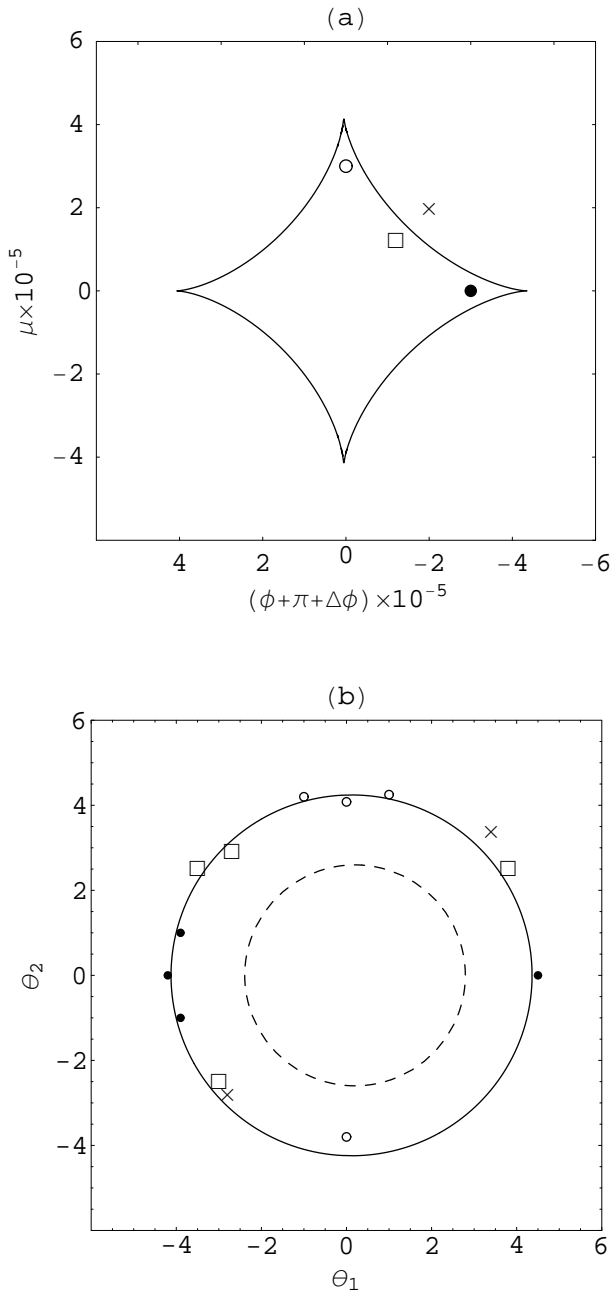}} \caption{A
schematic picture of the images of a point-source at $r_s=12M$
generated by a black hole with spin $a=0.2M$ as seen by an
equatorial observer. (a) The primary caustic cross-section and
four possible positions of the source, indicated by the cross, the
filled circle, the empty circle and the box. (b) The corresponding
images in the observer's sky together with the critical curve
(solid line) and the shadow of the black hole (dashed line).}
 \label{Fig Images}
\end{figure}

As anticipated in the introduction, the caustic separates two
regions of space in which a source gives rise to a different
number of images. After drawing the primary caustic, it is useful
to explain the way it affects the formation of the images around
the corresponding critical curve. We remind the reader that only
the images of order $m=1$ (i.e. generated by photons with one
polar inversion) are involved.

In Fig. \ref{Fig Images} we have drawn a schematic picture of the
images of a source close to the caustic as detected by an
observer. In Fig. \ref{Fig Images}a we have a cross-section of the
primary caustic at $r_s=12M$, along with several possible source
positions, represented by different symbols. In Fig. \ref{Fig
Images}b we show the corresponding critical curve as a solid line
and the shadow border in dashed style. The images corresponding to
the source positions of Fig. \ref{Fig Images}a are shown with the
same symbol. We also remind the reader that the spin is directed
toward the north pole ($\mu=1$) in Fig. \ref{Fig Images}a and is
parallel to the $\theta_2$ axis in Fig. \ref{Fig Images}b.

If the source is outside the caustic (the cross in Fig. \ref{Fig
Images}a) there are two images (the two crosses in Fig. \ref{Fig
Images}b). One image is on the same side of the source and lies
outside the critical curve (primary image) and the other is
opposite and lies inside the critical curve (secondary image). If
the source enters the caustic from the right (empty box), two
images with opposite parities are created on the left of the
critical curve (there is a reflection with respect to a vertical
axis from the caustic to the critical curve). If we place a source
on the equatorial plane close to the right cusp (filled circle),
we have three images close to the left tip of the critical curve,
corresponding to prograde photons. Finally, if we have a source
close to the upper cusp (empty circle), we have three images close
to the upper tip of the critical curve.

Then, to summarize, the right cusp of the caustic involves the
creation of images generated by prograde photons; the opposite
occurs for the left cusp. Therefore, we will often refer to the
right cusp as the prograde cusp and the left cusp as the
retrograde cusp. As for the upper and lower cusps, they are
associated with the formation of images generated by quasi-polar
orbits. In the primary caustic (and in all odd higher order
caustics), the upper cusp is associated with the upper tip of the
critical curve, whereas in all even order caustics it is
associated with the lower tip (there is also a reflection with
respect to a horizontal axis in even order caustics \cite{BDSS}).

This picture is a useful reference to understand how the different
parts of the caustic are associated with the the corresponding
parts of the critical curve and where the images are created or
destroyed when the source crosses a caustic. From now on, we will
concentrate mostly on the shape of the caustics, confining the
discussion about the critical curves to Section \ref{Sec Critics}.

\begin{figure}
\resizebox{\hsize}{!}{\includegraphics{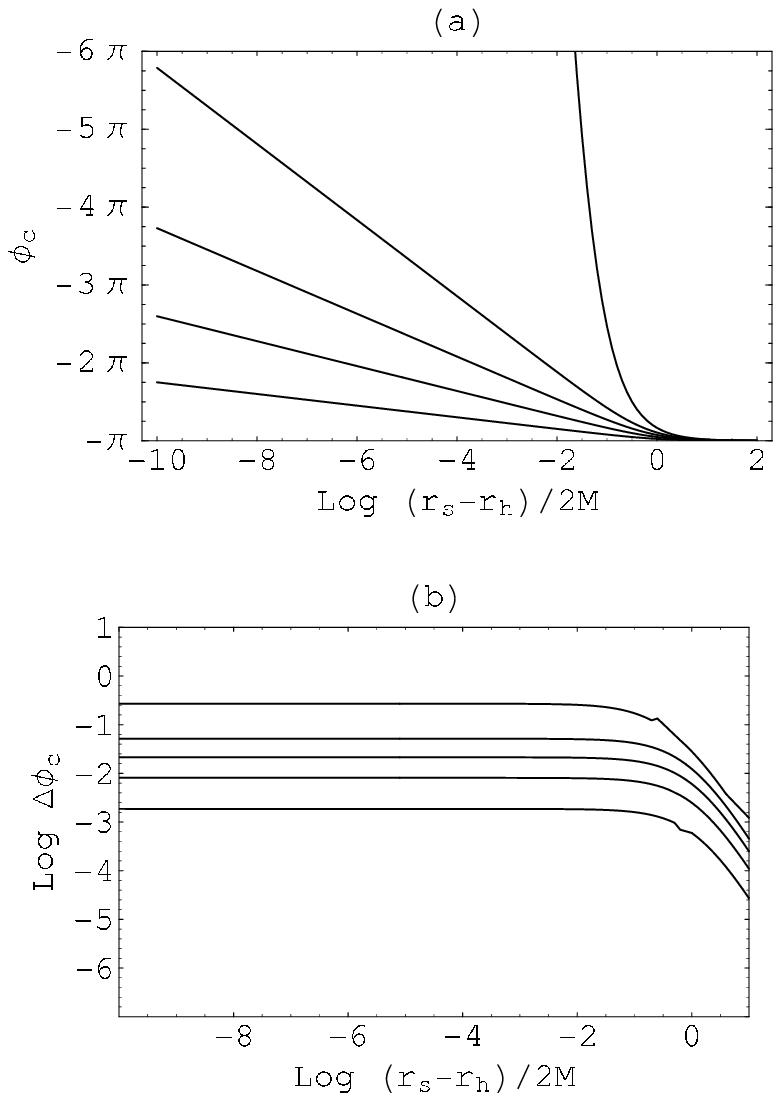}} \caption{(a)
Azimuthal coordinate of the retrograde cusp of the primary caustic
as a function of the radial coordinate. (b) Size of the primary
caustic as a function of the radial coordinate. In both figures,
from bottom to top, the curves are drawn for
$a=0.2M,0.4M,0.6M,0.8M,0.9998M$.}
 \label{Fig WFpossiz}
\end{figure}

Coming back to the primary caustic surface, we find that for any
values of the black hole spin the primary caustic performs an
infinite number of turns around the horizon. This fact is evident
from Fig. \ref{Fig WFpossiz}a, where the azimuthal coordinate of
the left cusp (corresponding to retrograde photons) is plotted vs
the logarithm of the difference between the radial coordinate and
the horizon radius $r_h$. This plot shows that at large distances
the primary caustic always tends to $-\pi$, whatever the value of
the spin. In the opposite limit, very close to the black hole
horizon, the position of the retrograde cusp shifts clockwise more
and more without stopping ($\phi$ increases with a negative sign).
We can easily establish a logarithmic law $\phi_c \simeq
c(a)\log(r_s-r_h)$, where the coefficient $c(a)$ depends on the
black hole spin, becoming larger and larger for high values of the
spin.

As already guessed in Ref. \cite{BozSca}, the divergence of the
azimuthal position of the caustic in the approach to the horizon
is a direct consequence of the divergence of the radial integral
$I_2$ (Eq. (\ref{I2}) in the appendix) appearing in the lens
equation for $\phi_s$ (\ref{Lens2}). This integral contains a
factor $1/\Delta$, which generates the logarithmic divergence of
$I_2$. This has been explicitly shown here by numerical
calculations for the first time.

In Fig. \ref{Fig WFpossiz}b we plot the size of the primary
caustic (calculated as the distance between the right and the left
cusp) as a function of the radial coordinate in logarithmic scale.
We see that the size tends to zero at very large distances,
whatever the value of the spin. This is only true for the primary
caustic, whereas all higher order caustics tend to a finite size
at large distances, as we shall see in the next section. Close to
the horizon, the size of the primary caustic tends to a constant.

\begin{figure}
\resizebox{\hsize}{!}{\includegraphics{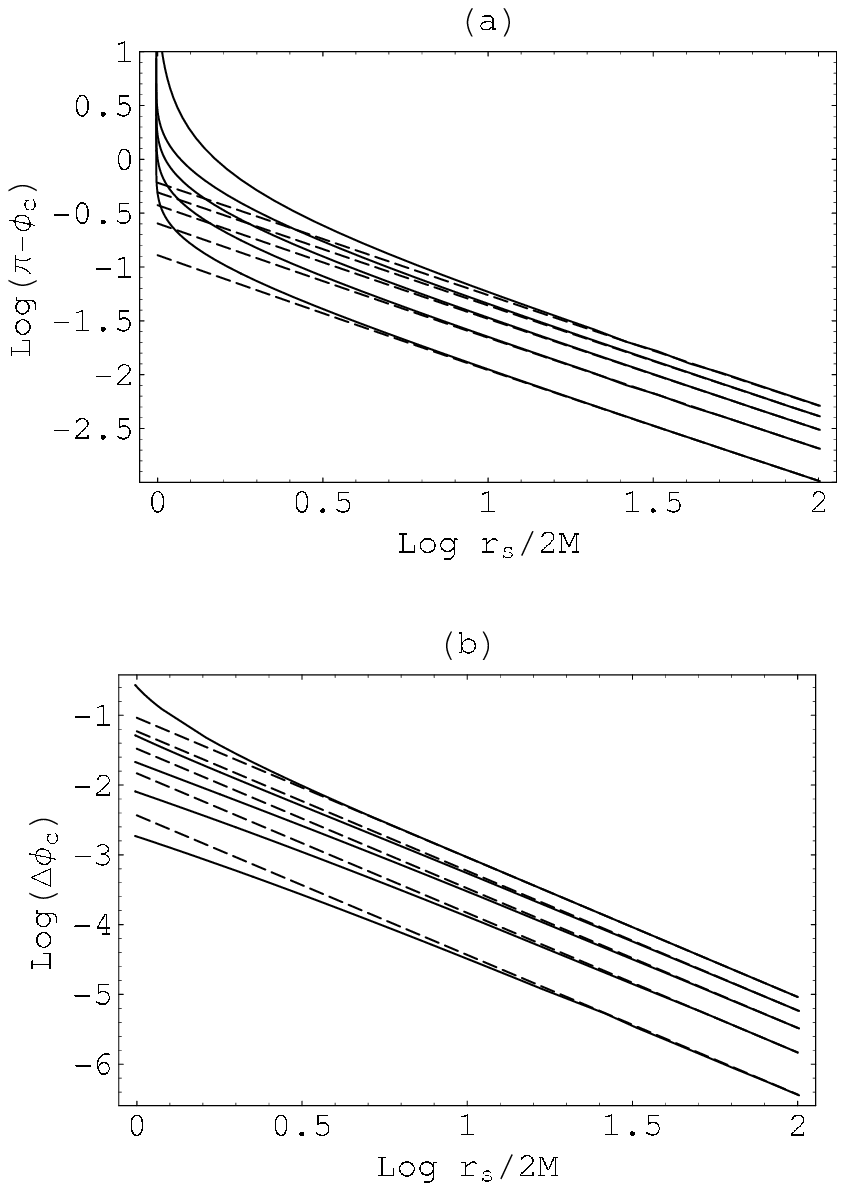}}
\caption{Comparison between numerical calculations (solid lines)
and the analytical approximation by Sereno and De Luca
\cite{SerDeL} (dashed lines). (a) Position of the retrograde cusp
as a function of the radial coordinate. (b) Size of the caustic as
a function of the radial coordinate. From bottom to top, the
curves are for $a=0.2M,0.4M,0.6M,0.8M,0.9998M$.}
 \label{Fig WFapp}
\end{figure}

Recently, an interesting analytical approximation for the primary
caustic has been proposed by Sereno and De Luca in Ref.
\cite{SerDeL}, using methods previously developed in Ref.
\cite{SerDeLa}. This approximation is obtained by an expansion in
the parameter
\begin{equation}
\varepsilon=\sqrt{\frac{M(r_o+r_s)}{4r_or_s}},
\end{equation}
which is valid at large distances from the black hole for any
values of the spin. The explicit expression of the caustic reads
\begin{eqnarray}
\phi_{s} & = & -\pi  -4 a \varepsilon^2 -\frac{5}{4}\pi a \varepsilon^3  \label{WFcaus1}  \\
 & + &
\left[ \left( \frac{225}{128}\pi^2 -16 \right) a -\frac{15}{16}
\pi a^2 \sqrt{1-\mu_o^2}
\cos^3 \eta  \right] \varepsilon^4 \nonumber \\
\mu_{s} & = & -\mu_o  -\frac{15}{16} \pi a^2
(1-\mu_o^2)^{3/2}\varepsilon^4 \sin^3 \eta  \label{WFcaus2},
\end{eqnarray}
with $\eta$ ranging from $-\pi$ to $\pi$.

In Fig. \ref{Fig WFapp} we compare our numerical results with the
formulae by Sereno and De Luca, finding a perfect agreement at
large distances from the black hole. In particular, Fig. \ref{Fig
WFapp}a represents the azimuthal position of the retrograde cusp
as a function of the radial coordinate in a log-log plot, whereas
Fig. \ref{Fig WFapp}b shows the size of the caustic as a function
of the radial coordinate in a log-log plot. The approximation only
fails at very small distances from the horizon. It is interesting
to note that the exact size is smaller than that predicted by Eq.
(\ref{WFcaus1}) for small spins, whereas the situation is reversed
at higher spins.

\section{Higher order caustics}

It is well-known that a Schwarzschild black hole generates two
infinite sequences of lensed images of a given source \cite{Dar}.
Apart from the two main images, usually treated in the Weak
Deflection Limit, all remaining images can be described in the
Strong Deflection Limit \cite{Dar,Cha,Boz1}. Higher order images
arise from the fact that photons grazing the unstable photon orbit
may perform one or more turns around the black hole before
emerging. For each turn, we have an additional pair of images,
whose luminosity decreases exponentially as we increase the number
of turns. In practice, gravitational lensing phenomenology is
replicated for each number of turns around the black hole,
including a new ring-like critical curve and a degenerate caustic
line. Furthermore, the structure is replicated even for sources in
front of the black hole (this geometrical configuration has been
dubbed retro-lensing \cite{Retro}). In this case, apart from the
direct image of the source, all other images are due to strongly
deflected photons turning around the black hole one or more times
and reaching back the observer.

Standard lensing and retro-lensing caustics can be treated in a
unified way, noting that we have a new caustic each time we add a
new inversion point in the polar motion of the photon. As
anticipated in Section II, the number of inversions can then be
identified with the caustic order. For example, the primary
caustic is generated by photons with a single inversion point, the
second order caustic (first retro-lensing caustic) is generated by
photons with two inversion points, the third order caustic (second
standard lensing caustic) is generated by photons with three
inversion points, and so on. This situation holds even when we
switch the angular momentum on. Nevertheless, analogously to what
happens for the primary caustic, higher order caustic tubes are
shifted and acquire a finite thickness. Up to now, higher order
caustics have been studied in the Strong Deflection Limit
approximation \cite{BozEq} in the equatorial plane and
successively for small values of the black hole spin in a fully
analytical way \cite{BDSS,BDS,BozSca}. In this section, we shall
present a complete study of higher order caustics in the whole
parameter space.

First we note that at large distances there is a fundamental
difference between the primary caustic and the higher order
caustics. In fact, whereas the primary caustic shrinks to zero
size and tends to the Schwarzschild point-caustic at $\phi=-\pi$
for $r_s \rightarrow \infty$, each higher order caustic tends to a
well-determined asymptotic astroid shape in the coordinates
$(\phi, \mu)$. Therefore, rather than a tube, at very large
distances the 3-dimensional region enclosed by the caustic
resembles a pyramid with astroidal base and vertex in the black
hole. The volume enclosed within the caustic grows as $ \Omega_c
r_s^2$, where $\Omega_c$ is the angular area within the asymptotic
astroidal shape.

\begin{figure}
\resizebox{\hsize}{!}{\includegraphics{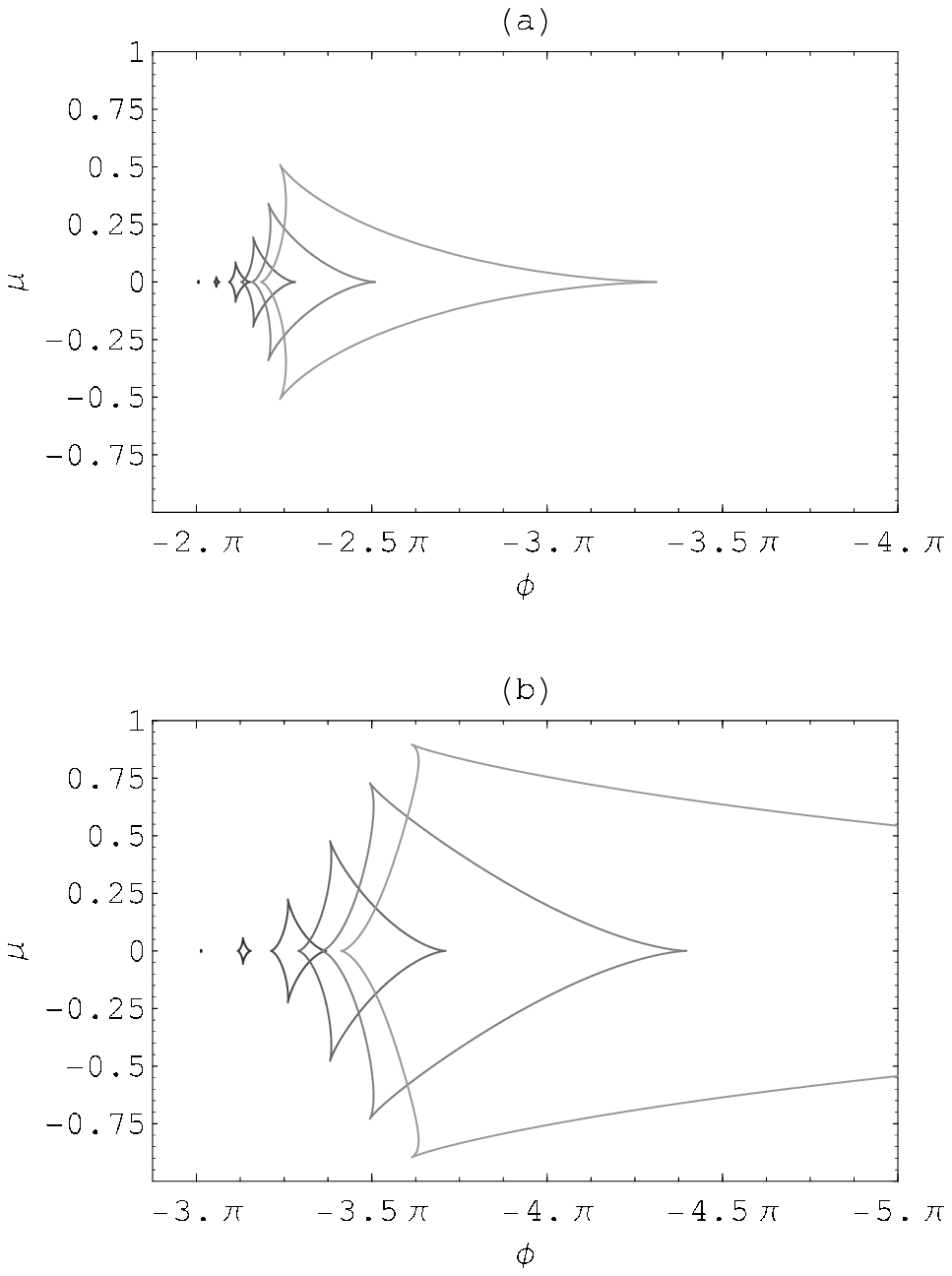}}
\caption{Second order (a) and third order asymptotic caustics (b)
for an equatorial observer $\mu_o=0$ and different values of the
spin. From left to right, $a=0.02M,0.2M,0.4M,0.6M,0.8M,0.9998M$.}
 \label{Fig a}
\end{figure}

Let us begin by studying the dependence of the asymptotic
cross-sections of the caustics at very large distances on the
black hole spin $a$. In Fig. \ref{Fig a} we show the second order
caustic (first retro-lensing caustic) and the third order caustic
(second standard lensing caustic) for different values of the
spin. At very low values of $a$, higher order caustics are very
small and only slightly displaced from the Schwarzschild positions
$\phi=-m\pi$. As $a$ grows, the size and the displacement of the
caustics grow. At the same time, the caustics become asymmetric,
with the cusp on the right (corresponding to prograde photons)
more stretched than the cusp on the left (retrograde photons).

At nearly extremal spins, higher order caustics become very large
but do not undergo any transitions to different shapes. This
regime is particularly interesting and will be deeply investigated
in the following section.

\begin{figure}
\resizebox{\hsize}{!}{\includegraphics{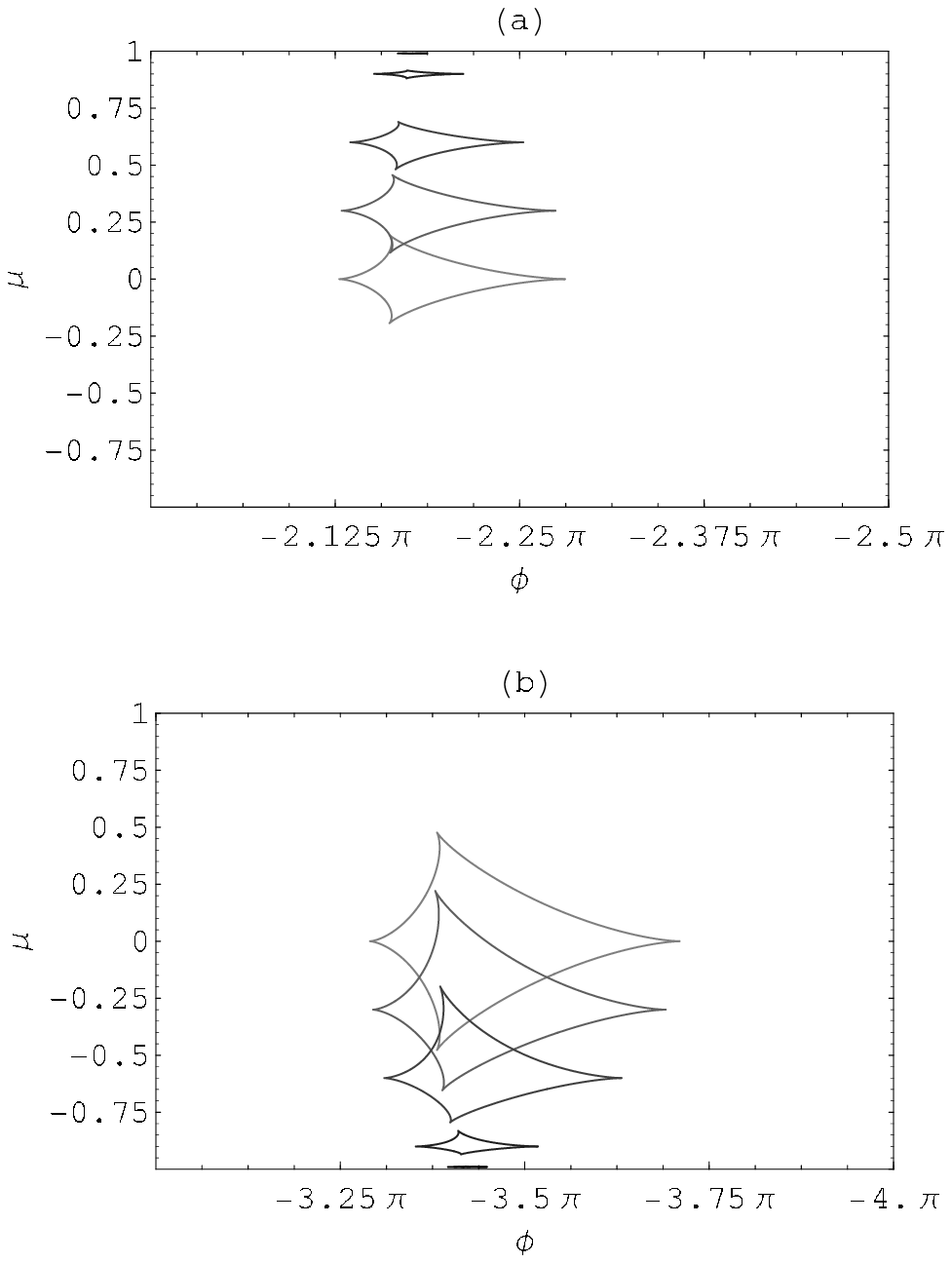}}
\caption{Second order (a) and third order asymptotic caustics (b)
generated by a Kerr black hole with spin $a=0.6M$ for different
values of the inclination $\mu_o=\cos \vartheta_o$. From the
center to the periphery, $\mu_o=0,0.3,0.6,0.9,0.99$.}
 \label{Fig muo}
\end{figure}

Now we turn to the dependence of the asymptotic caustics on the
spin inclination with respect to the line of sight. In Fig.
\ref{Fig muo} we show some asymptotic cross-sections of the
caustics for different values of the spin inclination $\mu_o$. The
retro-lensing caustics are always centered on $\mu_o$, i.e., they
lie at the same latitude as the observer, whereas the standard
lensing caustics lie at opposite latitude. Both retro-lensing and
standard lensing caustics preserve the same shift in the azimuthal
direction as the $\mu_o=0$ caustic. Increasing $\mu_o$ from $0$
(observer on the equatorial plane) to $1$ (observer on the polar
axis), all higher order caustics gradually shrink to zero size. An
observer on the polar axis would thus have point-like caustics on
the optical axis as in the Schwarzschild case. Another interesting
feature is that the caustics are slightly stretched on the side
closer to the equatorial plane. This is not just an effect of the
use of the projected coordinate $\mu$ as it can be seen even in
terms of the polar angle $\vartheta$. We must conclude that
multiple images arise more easily for sources close to the
equator.

In order to double-check our numerical calculations, we can
compare our asymptotic caustics with some approximation schemes.
The first is the one developed in Ref. \cite{BozEq}, based on the
Strong Deflection Limit. In that paper, only the intersections of
the caustics with the equatorial plane have been calculated
numerically. Nevertheless, a comparison with these results allows
us to check the retrograde and prograde cusp positions with a
completely independent calculation. The second approximation is
provided by the fully analytical formulae derived in Refs.
\cite{BDS,BDSS,BozSca} in the Strong Deflection Limit with an
expansion to second order in the black hole spin $a$. They read
\begin{eqnarray}
&& \mu_s= (-1)^m\left[\mu_o + R_m(1-{\mu_o}^2)^{3/2}\sin^3 \eta
\right],
\label{CauSDL1} \\
&& \phi_s=-m\pi-\Delta \phi_m+R_m\sqrt{1-{\mu_o}^2} \cos^3 \eta,
\label{CauSDL2}
\end{eqnarray}
where the azimuthal shift is
\begin{eqnarray}
&\Delta
\phi_m&=-\left\{\frac{2m\pi}{3\sqrt{3}}+2\log\left(2\sqrt{3}-3\right)
+ \right. \nonumber \\ && \left. \log\left[
\frac{(2\sqrt{r_{s}}+\sqrt{6M+r_{s}})}{3\sqrt{(r_{s}-2M)}}\right]
\right\} a, \label{Shift}
\end{eqnarray}
and the semi-amplitude of the caustic is
\begin{eqnarray}
&R_m&=a^2
 \left[\frac{1}{18} (5m\pi+8\sqrt{3}-36) \right. \nonumber \\
&& \left. +\frac{\left(
9M+2r_{s}-2\sqrt{r_{s}}\sqrt{6M+r_{s}}\right)}{3\sqrt{3}\sqrt{r_{s}}\sqrt{6M+r_{s}}}\right].
\label{Size}
\end{eqnarray}
In the limit $r_s \gg 2M$, these expressions reduce to their first
lines respectively.

\begin{figure}
\resizebox{\hsize}{!}{\includegraphics{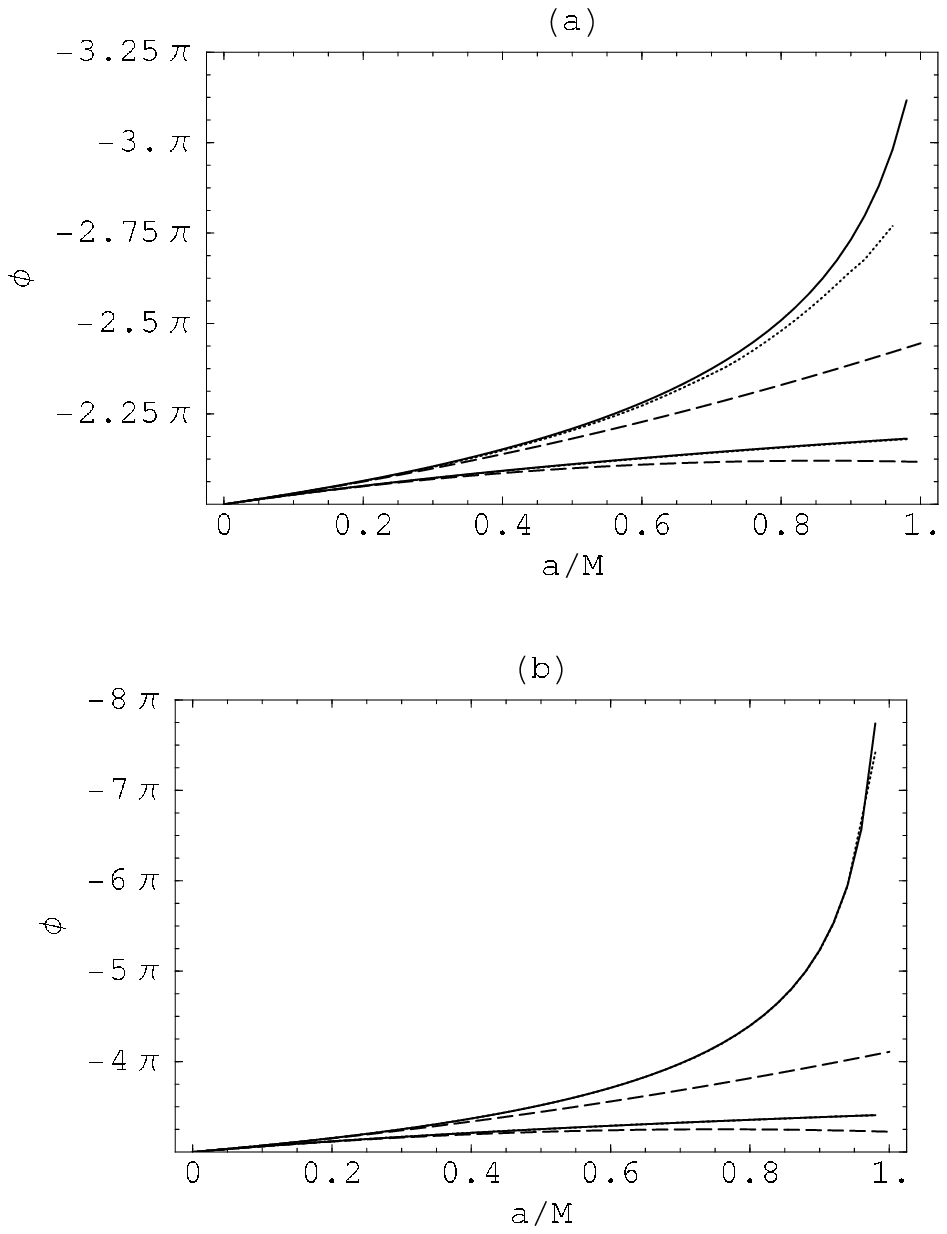}}
\caption{Azimuthal position of the prograde and retrograde cusps
of the asymptotic caustic as a function of the black hole spin.
(a) Second order caustic; (b) Third order caustic. The solid lines
are from numerical results in this paper, the dotted lines are the
SDL approximation of Ref. \cite{BozEq}, the dashed lines are the
perturbative results of Ref. \cite{BDS}.}
 \label{Fig SFcusp a}
\end{figure}

Fig. \ref{Fig SFcusp a} shows the dependence of the positions of
the cusps of the asymptotic caustic on the black hole spin. Along
with our numerical results, we have plotted the two approximations
described above. We see that the Schwarzschild limit is correctly
reproduced as $a\rightarrow 0$, since the caustics become
point-like and return to the optical axis ($-2\pi$ for the second
order caustic and $-3\pi$ for the third order caustic). As soon as
we switch the spin on, the caustics are shifted clockwise (more
negative $\phi$) and acquire finite extension, since the prograde
cusp is more shifted than the retrograde cusp.

The perturbative approximation works quite well up to $a\simeq
0.2M$. At higher spins, it underestimates the shift of the
caustics. On the other hand, the SDL approximation without the
expansion in powers of $a$ works extremely well for the third
order caustic, as the solid and dotted lines are practically
indistinguishable in Fig. \ref{Fig SFcusp a}b. For the second
order caustic, instead, it works quite well for the retrograde
cusp but fails at moderate and high spin values for the prograde
cusp. We must keep in mind that the SDL approximation is designed
for high caustic orders and is just marginally applicable to the
second order caustic.

\begin{figure}
\resizebox{5.86cm}{!}{\includegraphics{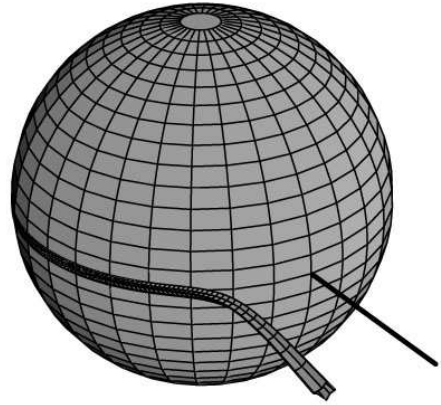}} \caption{The
second order caustic surface for $a=0.2M$ and $\mu_o=0$. The
straight line indicates the direction towards the observer.}
 \label{Fig SF3D}
\end{figure}

\begin{figure}
\resizebox{\hsize}{!}{\includegraphics{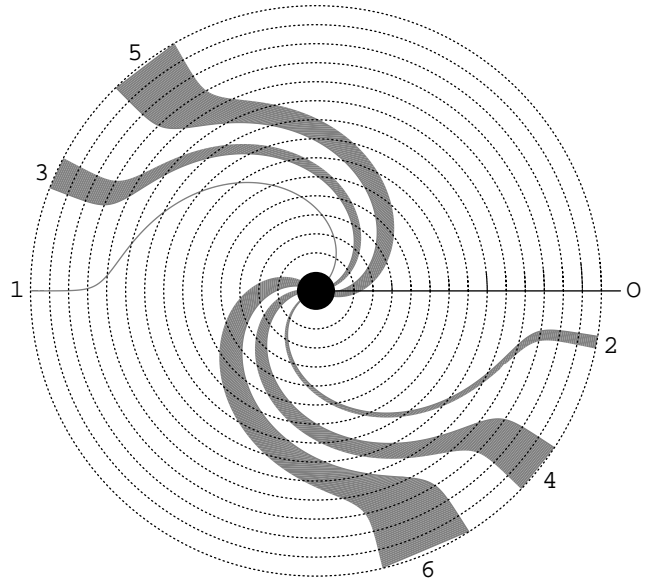}} \caption{A
representation of the caustics on the equatorial plane in which
the radial coordinate has been replaced by $\log (r_s-r_h)/2M+11$.
The outer circle represents the surface at $\log (r_s-r_h)/2M=4$.
Inner circle are in steps of 1 in such a logarithmic coordinate.
The observer is in the direction labelled by $O$ on the equatorial
plane. The number at the end of each caustic represents the
respective caustic order. Here the spin is $a=0.2M$ with the black
hole rotating counterclockwise.}
 \label{Fig spiral}
\end{figure}

Now let us come to the full development of the caustic surfaces,
from very far source distances up to the horizon. In Fig. \ref{Fig
SF3D} we show the second order caustic tube for a rather small
value of the spin ($a=0.2M$). It looks quite similar to the
primary caustic. The caustic tube has astroidal cross-section and
winds an infinite number of times around the horizon. However, the
cross-section does not shrink to zero size at infinity, but
preserves the same angular size. A 3-dimensional picture
containing several caustics at the same time would not be so
readable, as all caustics wind around the horizon so tightly that
they would be undistinguishable. However, a picture giving the
idea of how several caustics wind around the black hole is in Fig.
\ref{Fig spiral}, where we have represented the projection of the
caustics on the equatorial plane. The radial coordinate has been
replaced by the logarithmic coordinate $\log (r_s-r_h)/2M+11$ in
order to put the ``spiral arms'' of the caustics in better
evidence. Of course, such a representation becomes meaningless at
distances less than $r_h+10^{-11}M$. We have represented the
primary caustic (labelled by ``1") along with higher order
caustics up to the sixth order. In this way we have explicitly
shown where the caustics lie with respect to the observer's
direction (labelled by ``$O$''). In the Schwarzschild limit, all
caustics reduce to a line: odd ones are on the opposite side and
even ones (retro-lensing caustics) are on the same side as the
observer. At $a>0$, all caustics are shifted clockwise and acquire
a finite extension, proportionally to their order.

\begin{figure}
\resizebox{\hsize}{!}{\includegraphics{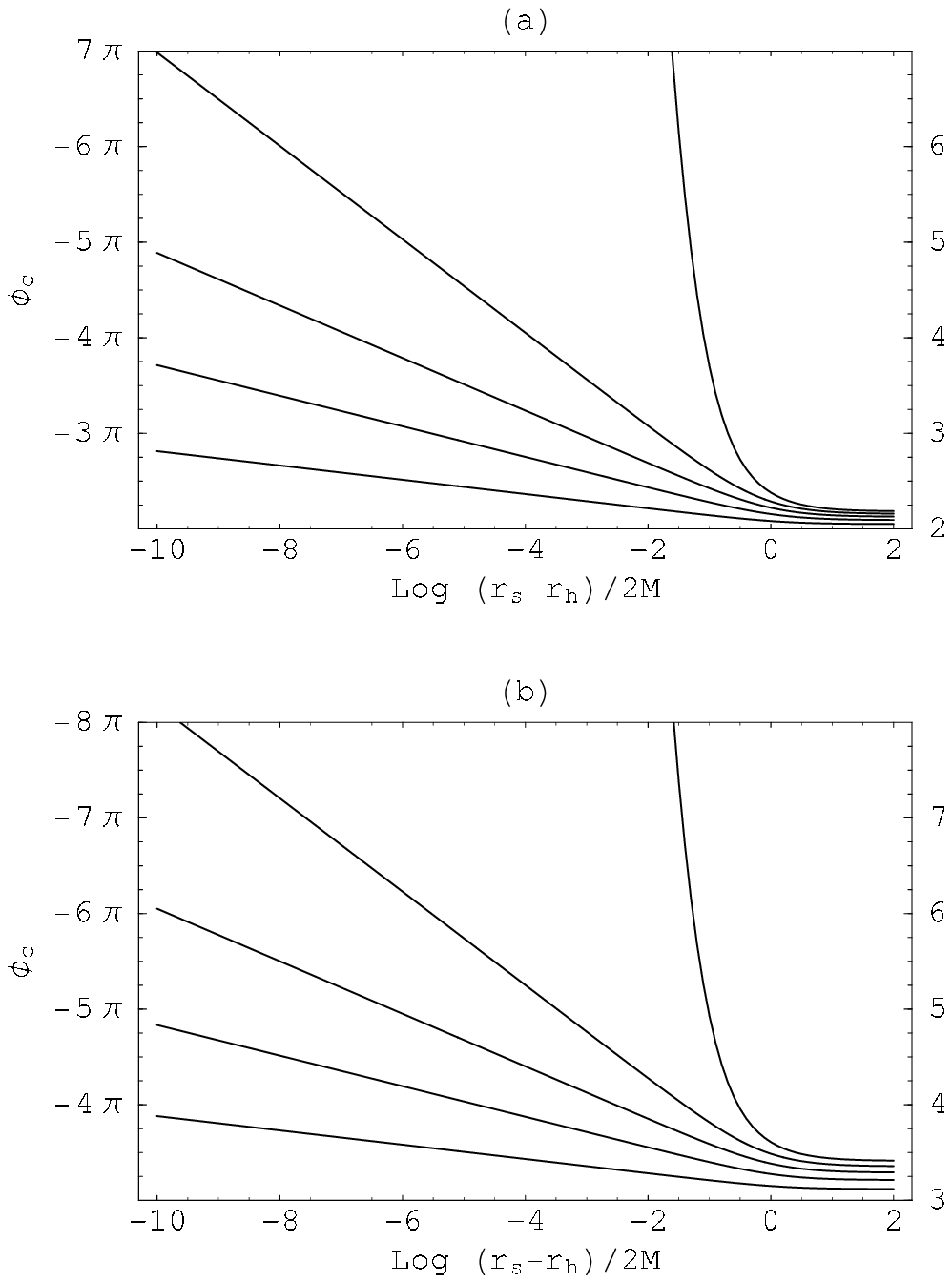}}
\caption{Position of the retrograde cusp for the second order (a)
and the third order (b) caustics, as a function of the radial
coordinate. In both figures, from bottom to top, the curves are
for $a=0.2M,0.4M,0.6M,0.8M,0.9998M$.}
 \label{Fig SFcuspos}
\end{figure}

\begin{figure}
\resizebox{\hsize}{!}{\includegraphics{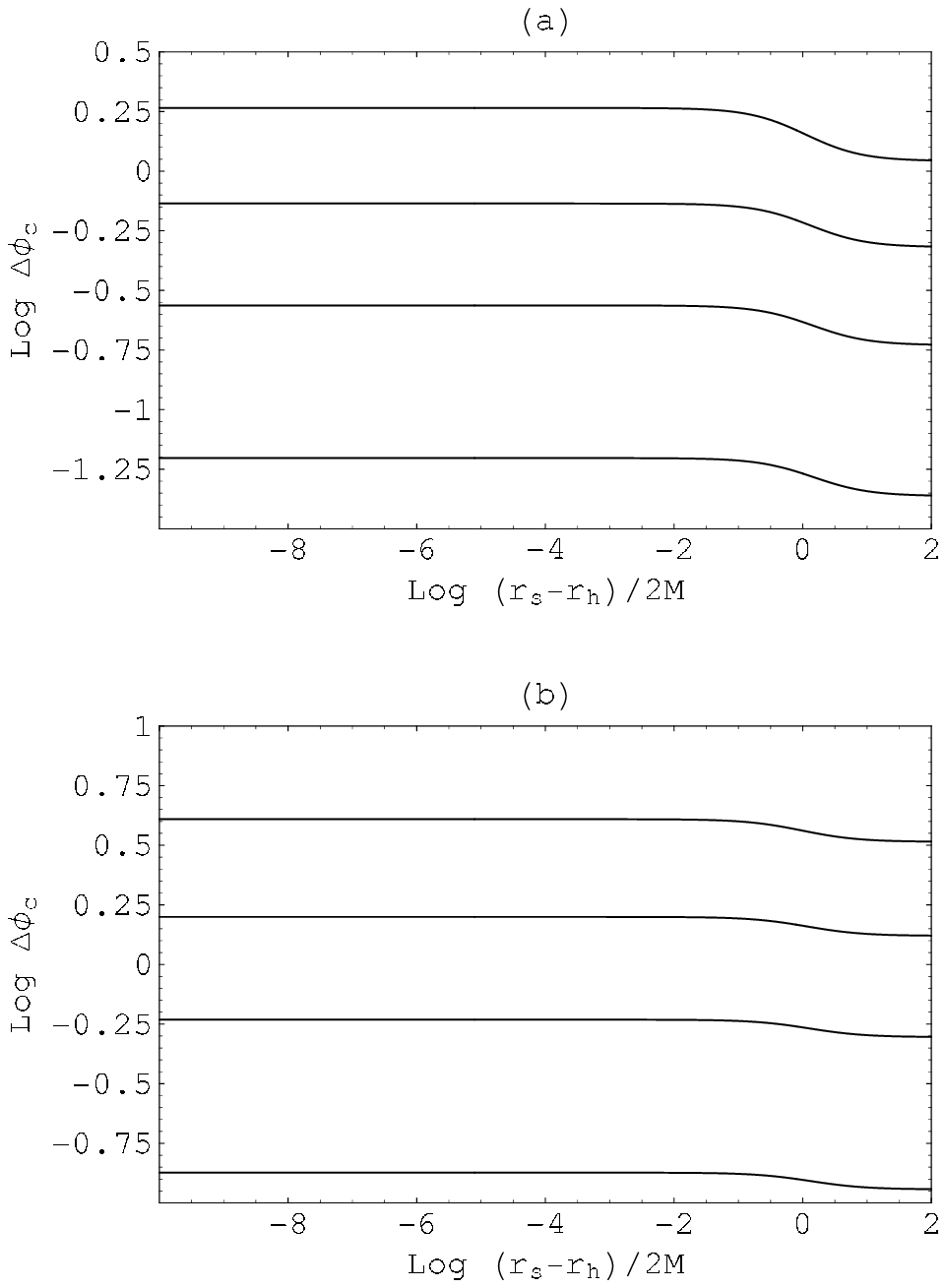}}
\caption{Size of the second order (a) and the third order (b)
caustics, as a function of the radial coordinate. In both figures,
from bottom to top, the curves are for $a=0.2M,0.4M,0.6M,0.8M$.}
 \label{Fig SFcussiz}
\end{figure}

Let us come to a more quantitative analysis. Fig. \ref{Fig
SFcuspos} shows the azimuthal position of the retrograde cusp as a
function of the radial coordinate for the second and third order
caustic. The behavior is very similar to the primary caustic, with
the logarithmic divergence for $r_s\rightarrow r_h$. However, at
large radii, the retrograde cusp does not settle to the
Schwarzschild position $-m\pi$, but a finite shift remains. In
Fig. \ref{Fig SFcussiz} we plot the size of the caustics
(estimated as the distance between the prograde and the retrograde
cusp) as a function of the radial coordinate. Indeed, the size of
the caustics tends to two different constants at large and small
radii. This behavior has already been noted for small $a$ caustics
in Ref. \cite{BozSca}.

\begin{figure}
\resizebox{\hsize}{!}{\includegraphics{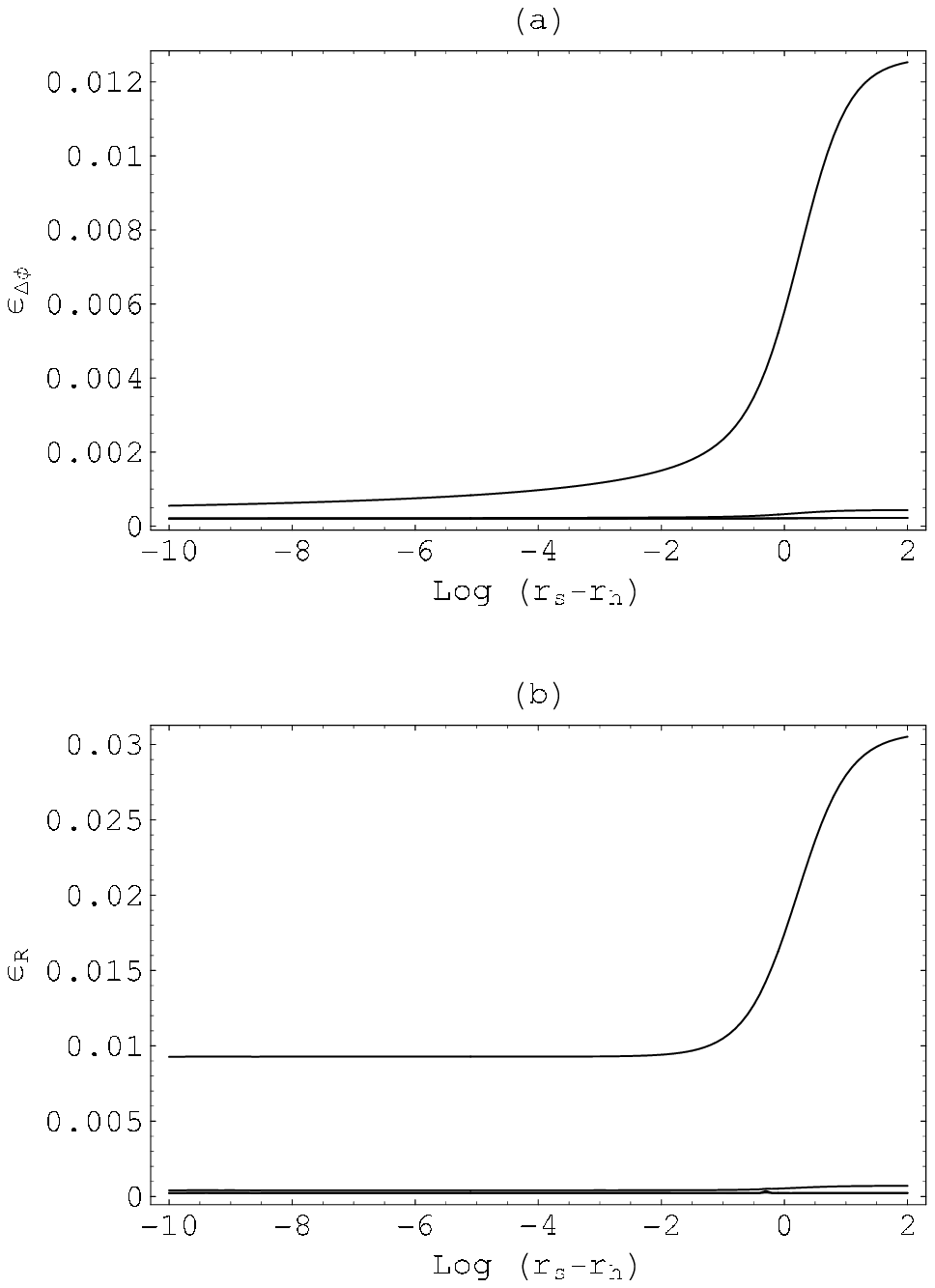}}
\caption{Relative error of the perturbative approximation given by
Eqs. (\ref{Shift}) and (\ref{Size}) as function of the radial
coordinate. (a) Error in the shift of the caustic (b) Error in the
size of the caustic. Curves are for caustic orders $m=2,3,4$ from
top to bottom. $a=0.02M$, $\mu_o=0$.}
 \label{Fig SFapp}
\end{figure}

We conclude this section by a picture representing the relative
error in the prediction of the position and size of the caustics
by the perturbative formulae (\ref{Shift}) and (\ref{Size}). In
Fig. \ref{Fig SFapp} we can see that the error decreases
exponentially with the caustic order $m$, as predicted by the SDL
approximation. However, a residual error remains because of the
expansion in powers of $a$. In any case, for the value $a=0.02M$
used in this plot, the error is at most of $3\%$ for the second
order caustic at large distances. Therefore, the SDL approximation
works amazingly well at all radial distances, reproducing the
logarithmic spiralling perfectly.

\section{Caustics of extremal Kerr black holes}

As shown in the previous sections, higher order caustics become
very large at high values of the spin. Since there are several
indications that nearly extremal spinning black holes may be not
uncommon in the universe, it is particularly important to focus on
the caustic structure of this particular limiting species of Kerr
black holes.

\begin{figure}
\resizebox{\hsize}{!}{\includegraphics{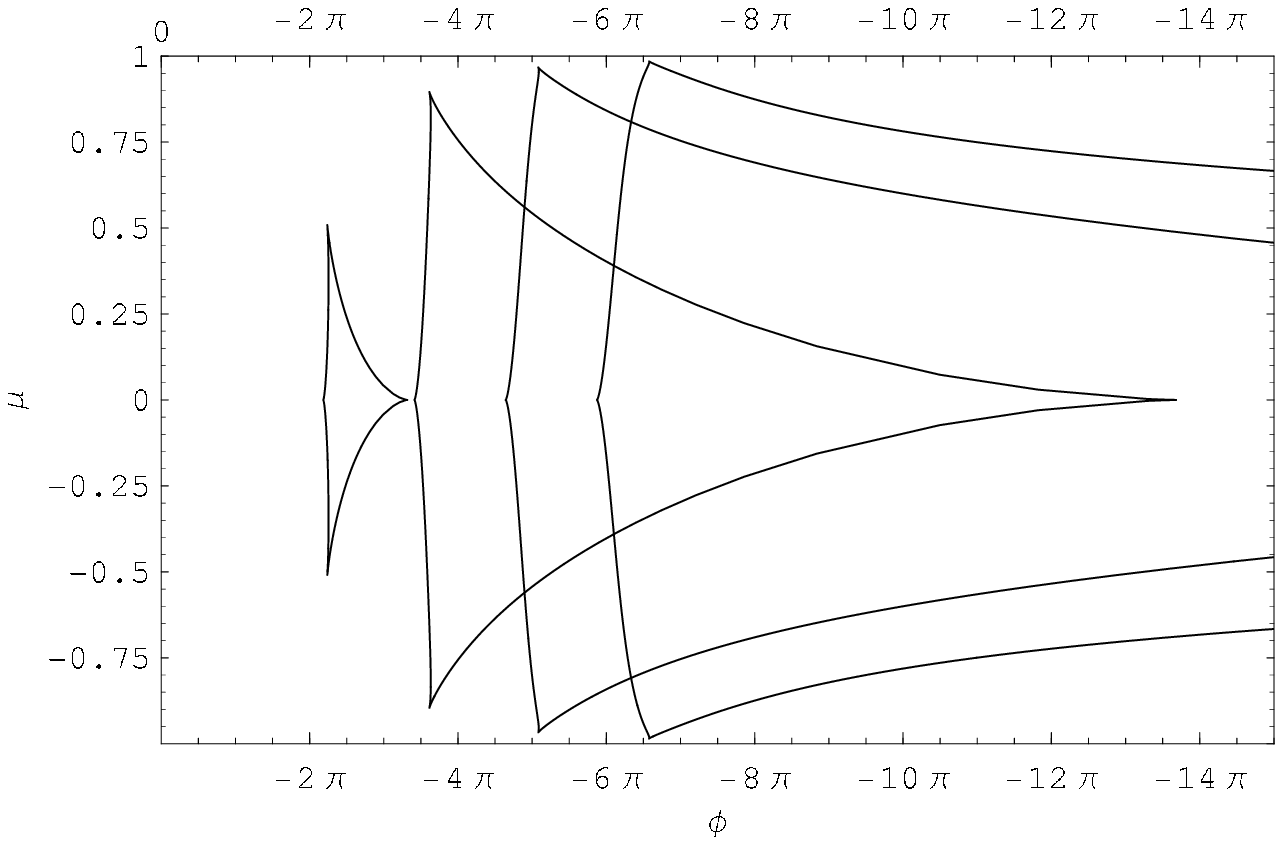}}
\caption{Asymptotic caustics of an extremal Kerr black hole for an
equatorial observer. From left to right the caustic order is
$m=2,3,4,5$.}
 \label{Fig ExtOrders}
\end{figure}

First, we show the asymptotic cross-sections at large source
distances for several caustic orders. Apart from the second order
caustic, all higher order caustics extend for more than a complete
loop around the black hole. The left (retrograde cusp) is shifted
clockwise from the Schwarzschild position linearly with the
caustic order $m$, following the law
\begin{equation}
\phi_l=-3.86m+0.87. \label{phil}
\end{equation}
On the contrary, the right (prograde) cusp is shifted
exponentially with $m$, following the law
\begin{equation}
\phi_r=-1.05m-0.36 \exp (1.57m).  \label{phir}
\end{equation}
Caustics overlap each other without any interference, as they
correspond to trajectories with different numbers of inversion
points in the polar motion, which live in different regions of the
($\psi$,$\eta$) space. Also the extension in latitude increases,
with the northern and southern cusps gradually approaching the
poles.

\begin{figure}
\resizebox{\hsize}{!}{\includegraphics{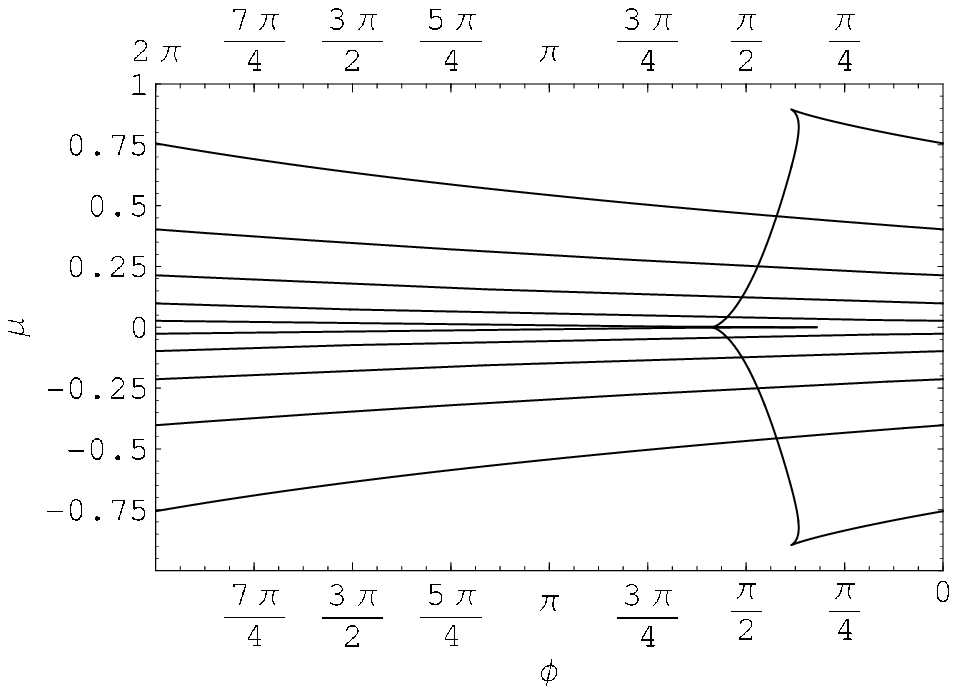}} \caption{Third
order asymptotic caustic of an extremal Kerr black hole for an
equatorial observer in the $\phi$-$\mu$ plane. In this plot, we
have taken into account the fact that the azimuthal coordinate is
periodic with period $2\pi$.}
 \label{Fig Ext3rd}
\end{figure}

The reader may wonder about the physical consequences of having
caustics extending for more than $2\pi$ in azimuth. In Fig.
\ref{Fig Ext3rd} we plot the third order caustic reporting all
points in the physical interval $0\leq \phi< 2\pi$. In this way,
the caustic appears to exit from the right side and re-enter from
the left side for five times before it ends in the prograde cusp.
Suppose to have a source on the same side of the observer ($\phi_s
\simeq 0$) and close enough to the northern pole ($\mu_s \simeq
1$) to be outside of the caustic surface. In this situation, there
are only two images formed by photons with three inversions in the
polar motion. If we move the source down, by decreasing $\mu_s$,
the source meets the caustic for the first time. At this point,
two new images appear, as we know from gravitational lensing
theory. Their initial position on the critical curve is the one
that corresponds to the caustic point where the source crosses the
caustic. However, if we continue to move the source down, at some
point we meet the caustic for the second time. Two new images
appear in a different position on the critical curve. In
particular, these new images are still formed by photons with
three inversion points in the polar motion but they perform one
more loop in the azimuthal motion before reaching the observer
with respect to the previous pair. If we continue to decrease
$\mu_s$, we can cross the caustic 4 more times, including the
final cusp crossing. Therefore, for an extremal Kerr black hole,
we can have up to 14 images with three inversion points in the
polar motion.

The number of times a particular caustic can be crossed is simply
given by the number of azimuthal loops spanned by the caustic plus
one. With the empirical formulae (\ref{phil}) and (\ref{phir}), we
can estimate that a caustic of order $m$ may generate up to
\begin{equation}
n=[0.36 \exp (1.57m)-2.82m+0.87]
\end{equation}
new pairs of images. In practice, for a source very close to the
pole, the number of images with $m$ inversions in the polar motion
will always be the minimal one, i.e. 2. On the other hand, for a
source close to the equatorial plane, the number of images will
always be very close to the maximal one $2(n+1)$. It is
interesting to connect this result with the exponential decrease
in the magnification of higher order images with the caustic
order. From one side we have an exponential increase of the number
of images. From the other side we have an exponential decrease in
the magnification factor of each image. However, even if we have
not calculated the details of the magnification, we can expect
that the decrease in the magnification dominates on the increase
in the number of images, in such a way that the total flux in the
images of order $m$ still decreases exponentially with the caustic
order. Finally, we also note that, apart from the first pair of
additional images, all images created by repeated caustic
crossings occur in a region close to the stretched prograde cusp.
Therefore, all these additional images will appear to the observer
on the left side of the black hole.

\begin{figure}
\resizebox{\hsize}{!}{\includegraphics{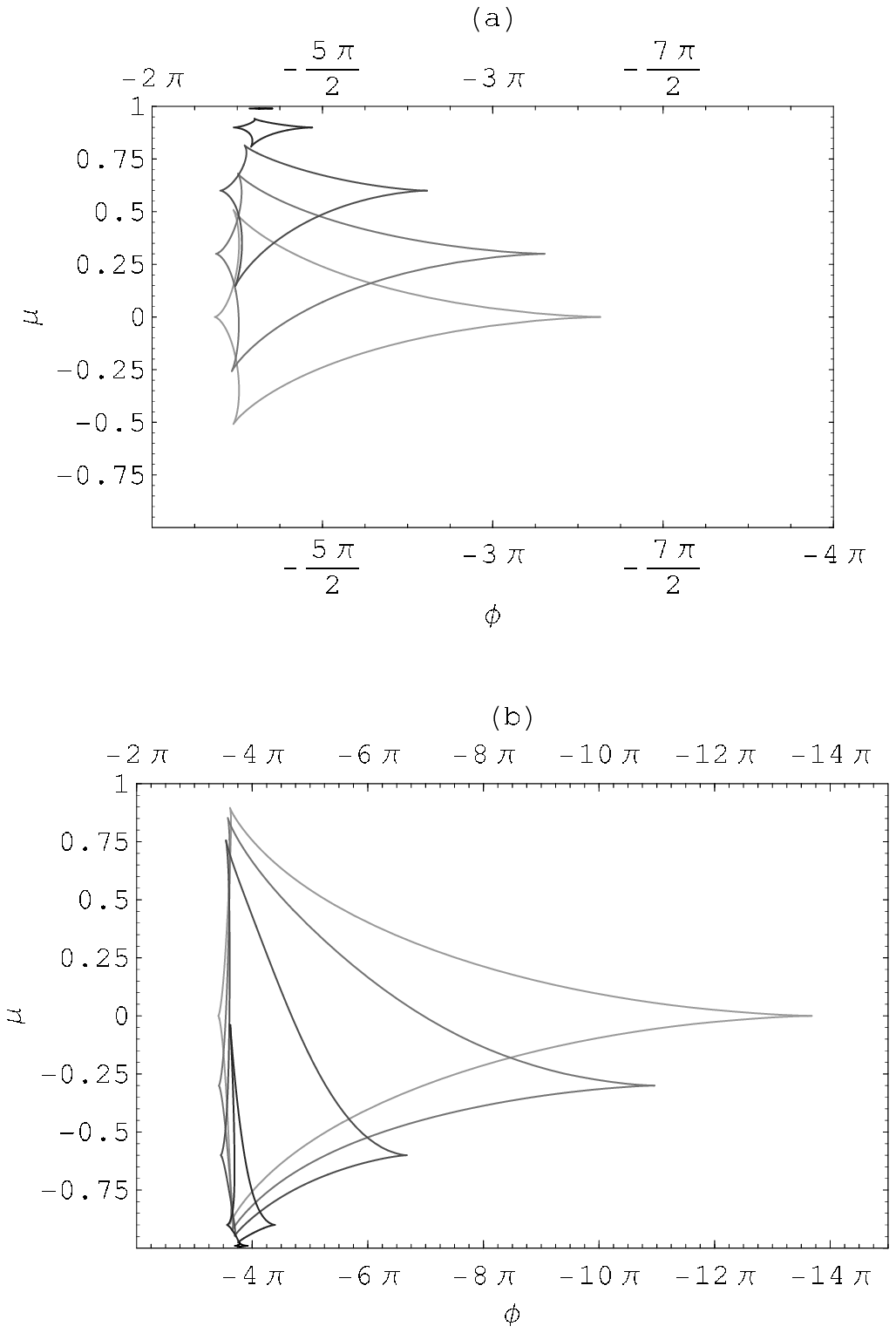}}
\caption{Second order (a) and third order (b) asymptotic caustics
of an extremal Kerr black hole for different positions of the
observer. From the center to the periphery the caustics are drawn
for $\mu_o=0,0.3, 0.6,0.9,0.99$.}
 \label{Fig Extmuo}
\end{figure}

\begin{figure}
\resizebox{5.86cm}{!}{\includegraphics{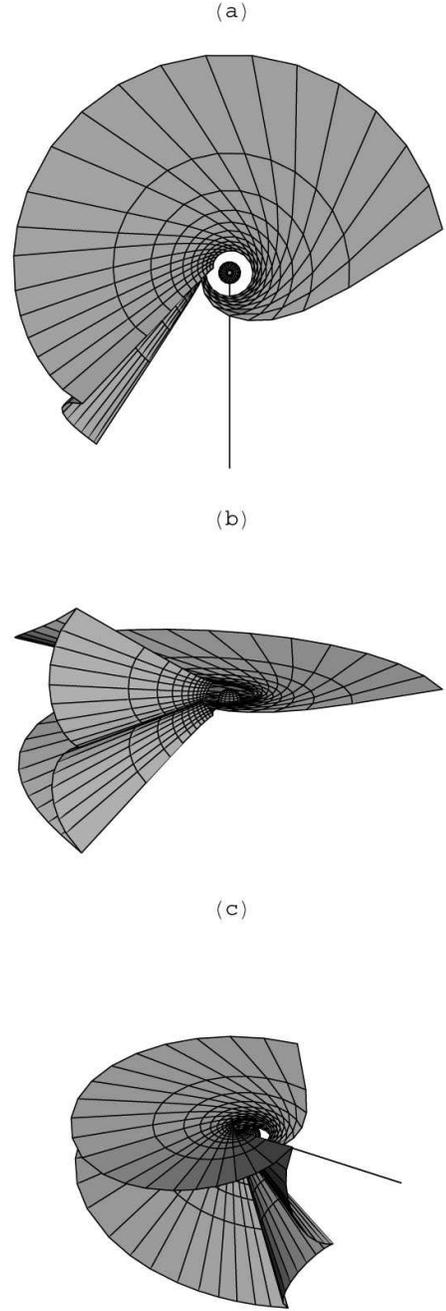}}
\caption{3-dimensional pictures of the second order caustic
surface of an extremal Kerr black hole for an observer on the
equatorial plane. The surface has been plotted for radial
distances in the range $[2.2M,20M]$. (a) View from the north pole.
(b) View from the observer side, lifted by $10^\circ$ above the
equator. (c) View from the left side. The straight line in (a) and
(c) points towards the observer.}
 \label{Fig Ext3D}
\end{figure}

As for the dependence of the caustics on the spin inclination on
the line of sight, we have plotted several second order and third
order caustics in Fig. \ref{Fig Extmuo}. As in Fig. \ref{Fig muo},
the caustics keep the same azimuthal shift for their central
region while their size decreases with increasing $\mu_o$. Even if
the $\mu_o=0$ caustics extend for several azimuthal loops, by
increasing $\mu_o$ up to $\mu_o=1$, all caustics shrink to a
point. Furthermore, we can see in particular from Fig. \ref{Fig
Extmuo}b that the caustics maintain their extension in latitude
until the azimuthal stretch is reduced to less than one loop. Only
then, the latitudinal size starts to decrease significantly. As a
consequence, since the left and right cusps are always at $\mu=\pm
\mu_o$, we have a considerable asymmetry between the side of the
caustic towards the pole and the side towards the equator, which
remains much larger.

\begin{figure}
\resizebox{\hsize}{!}{\includegraphics{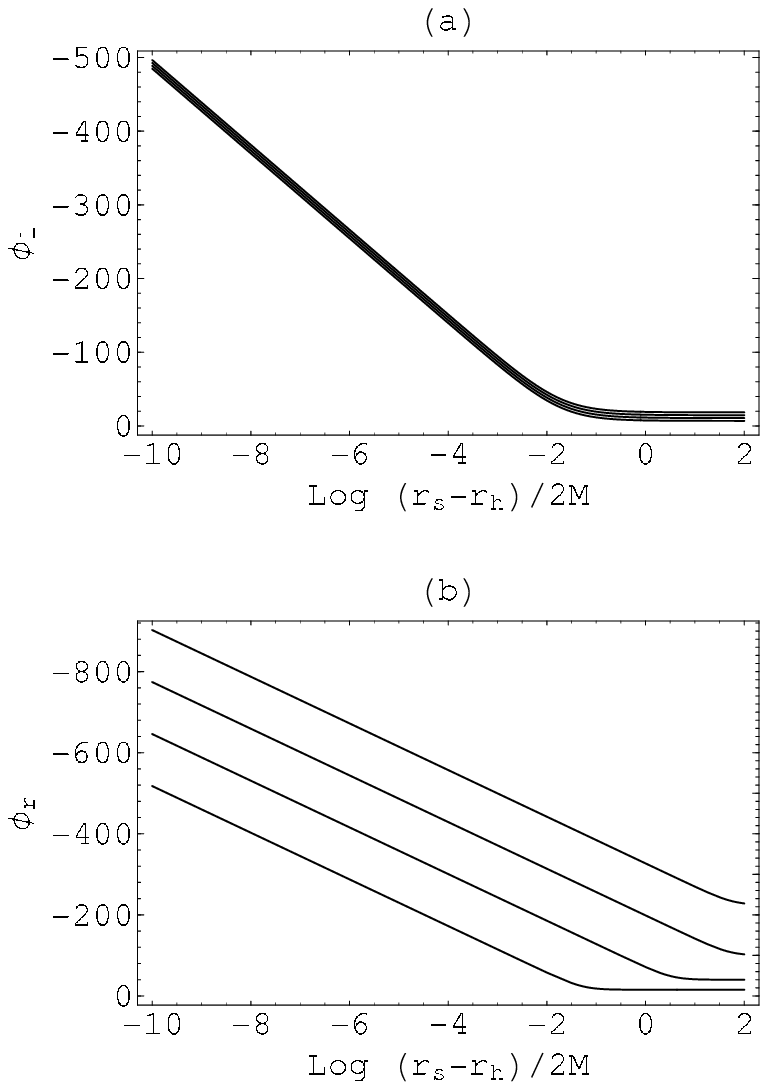}}
\caption{Azimuthal position of the retrograde cusp (a) and
prograde cusp (b) as a function of the logarithm of the distance
from the event horizon. From bottom to top, the curves are for
caustic order $m=2,3,4,5$.}
 \label{Fig Extcuspos}
\end{figure}

After the study of the asymptotic cross-section of the caustic
surfaces, it is time to examine the full development of the
caustics of an extremal Kerr black hole. As it can be deduced from
the extension of the asymptotic cross-sections, the caustic
surfaces become so wide that they no longer resemble a tube. They
assume an almost disk-like shape with the thickness increasing
with the distance from the black hole. Some 3-dimensional pictures
of the second order caustic are shown in Fig. \ref{Fig Ext3D} with
different points of view. The caustic has been calculated for an
observer on the equatorial plane (whose direction is indicated by
the straight line). One can appreciate the half-disk shape which
winds around the black hole more and more. The retrograde cusp is
the equatorial cusp on the left side, which also enjoys the wider
latitudinal extension. The prograde cusp is on the right side,
where the caustic has a spiky cross-section. The prograde cusp
winds around the black hole much more quickly than the retrograde
cusp, at least initially. It can be easily imagined that the
pictures of even higher order caustic surfaces, which spread over
several loops around the black hole, are very difficult to show in
a clear way. The caustic surface self-intersects several times.

Finally, we plot the position of the retrograde and prograde cusps
in Fig. \ref{Fig Extcuspos}. We can see that both retrograde and
prograde cusps wind around the black hole following the usual
logarithmic law $\phi\sim c \log (r_s/r_h-1)$. However, the
prograde cusps start to follow the logarithmic law at much larger
distances than the retrograde cusps. This means that the caustic
has a very large increase from the asymptotic size at large radial
distances to the final size at distances very close to the black
hole. The slope of the logarithmic laws is the same for all
prograde and retrograde cusps and amounts to $c=57.55$, which
means that each time we decrease the distance from the horizon by
a factor 10, the caustic makes $9.16$ turns around the black hole.

\begin{figure*}
\resizebox{\hsize}{!}{\includegraphics{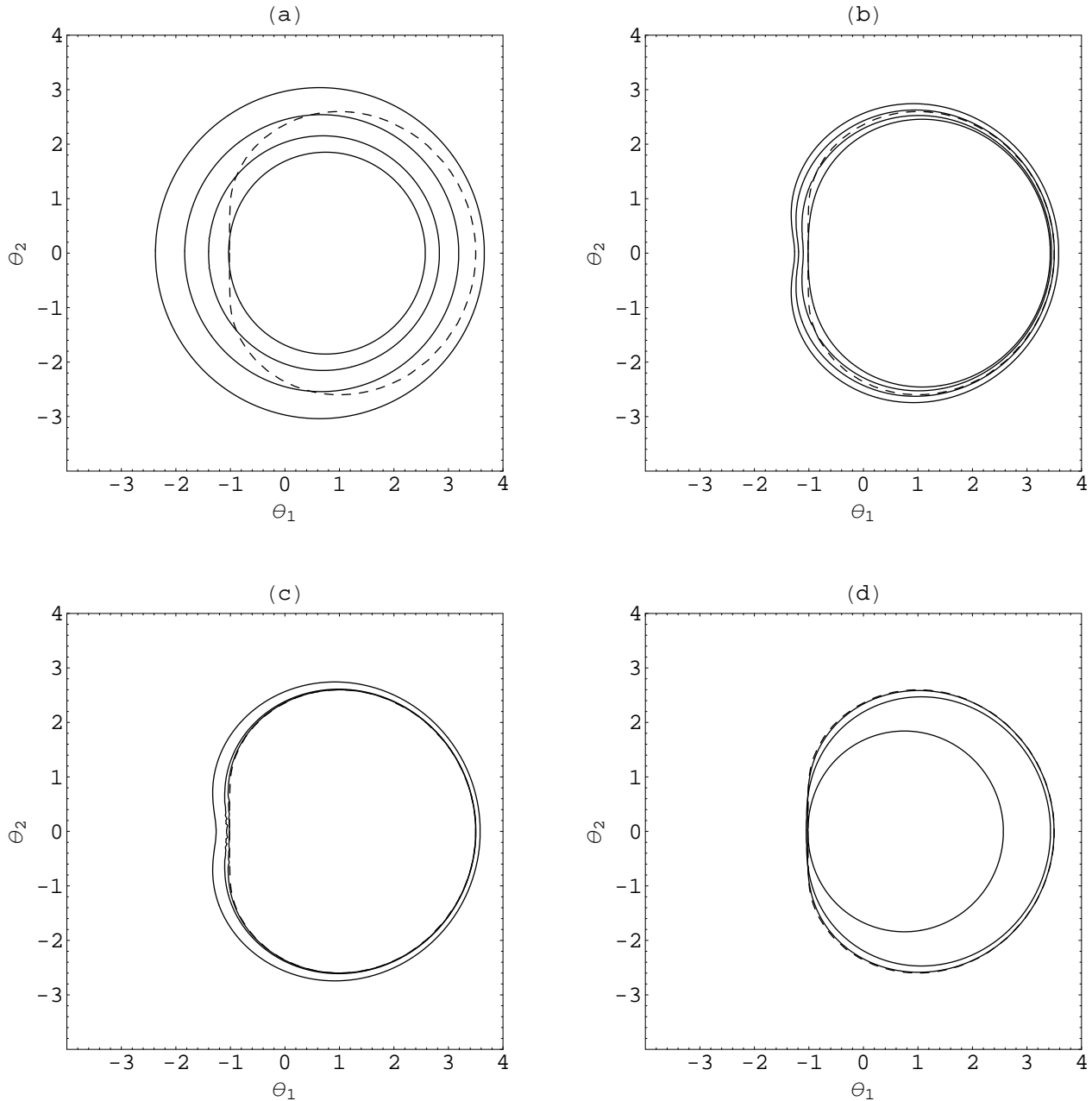}}
\caption{Critical curves for a Kerr black hole with spin
$a=0.9998M$ and an equatorial observer. The coordinates
$(\theta_1,\theta_2)$ in the observer's sky are in units of
$2M/r_o$. (a) Primary critical curve for various source distances:
starting from the outside, $r_s=5M,3M,1.8M,1.1M$. (b) Second order
critical curve for various source distances: starting from the
outside, $r_s=\infty,4.2M,1.6M,1.1M$. (c) Critical curves of order
$m=2,3,4$ for a source at infinity. (d) Critical curves of order
$m=1,2,3$ for a source at $r_s=1.1M$. The shadow border is
represented by the dashed curve.}
 \label{Fig Critics}
\end{figure*}

\section{Critical curves} \label{Sec Critics}

As stated in Section II, once we find a critical point of the lens
mapping in the $(\psi,\eta)$ space, we can easily determine the
corresponding critical point in the observer's sky, spanned by the
angular coordinates $(\theta_1,\theta_2)$. As $\eta$ varies from
$-\pi$ to $\pi$, the critical points describe a closed curve
around the black hole. Therefore, for each caustic order $m$, for
a given source distance $r_s$, we have a distinct critical curve
in the observer's sky. In particular, the primary critical curve
of order $m=1$ is the classical Einstein ring. If a source at
distance $r_s$ lies on the primary caustic tube and is large
enough that a complete cross-section of the caustic tube lies
inside the source, then the observer will see the primary image
and the secondary image merged together to cover the whole
Einstein ring. Since higher order caustics have a greater
extension, the case that a source intercepts a complete
cross-section of the caustic tube can only occur for sufficiently
low spins or very large sources. Therefore, it is much more
difficult to have a complete Einstein ring of higher order.
Nevertheless, critical curves are the loci where additional pairs
of images appear or disappear when a caustic crossing occurs.
Since such images are maximally magnified during the process of
creation or destruction, it is important to study critical curves
because they indicate the places where higher order images are
most likely to be detected. However, the average radius of a
critical curve depends on the source distance. This is
particularly evident for the primary critical curve, whose average
radius at large source distances scales as the Einstein angle
$\theta_E =\sqrt{4Mr_s/r_o(r_o+r_s)}$. Higher order critical
curves tend to an asymptotic shape for large source distances,
whereas they become smaller for smaller source distances.
Typically, higher order critical curves are slightly larger than
the shadow of the black hole, but when the source is closer than
the unstable photon orbits, the critical curves appear entirely
inside the shadow of the black hole.

In Fig. \ref{Fig Critics} we show some critical curves for a
nearly extremal Kerr black hole and an equatorial observer. As
well-known \cite{Cha}, the shadow is shifted to the right
(retrograde side) with respect to the black hole position, with
the left (prograde) side more flattened.

In Fig. \ref{Fig Critics}a we show the primary critical curve for
various source distances. If the source is far from the black
hole, the critical curve is also far from the shadow. If the
source is closer to the black hole, the critical curve shrinks
gradually and finally enters the shadow when the source is closer
to the black hole. Note that the critical curve enters the shadow
from the retrograde side first, because the retrograde unstable
orbit lies at a larger radius. The prograde unstable orbit is
closer to the horizon and therefore the critical curve stays
outside the shadow on the prograde side until the source becomes
very close to the horizon. We can also note that the primary
critical curve maintains its nearly circular shape at all source
distances.

Higher order critical curves, instead, follow the shape of the
shadow more closely. In Fig. \ref{Fig Critics}b we show the second
order critical curve for various source distances. At very large
source distances each critical curve tends to a fixed asymptotic
curve at finite distance from the shadow. As the source distance
is decreased, the critical curve shrinks and finally enters the
shadow. Note that higher order critical curves have a small
concavity on the prograde side, which disappears when the source
is closer to the black hole.

In Fig. \ref{Fig Critics}c we show the critical curves of order
$m=2,3,4$ altogether for a source at infinity. All critical curves
are outside the shadow with higher order ones closer to the
shadow. In Fig. \ref{Fig Critics}d we show the critical curves of
order $m=1,2,3$ for a source at $r_s=1.1M$ i.e. very close to the
horizon. In this case the primary curve is the most internal one,
with the higher order ones gradually closer to the shadow.

Finally, the dependence of the critical curves on the spin
inclination closely follows the dependence of the shadow, which
becomes more and more circular and symmetric as the observer's
latitude increases.

\section{Conclusions}

In all applications of gravitational lensing, the study of
critical curves and caustics of specific lens models has always
represented a fundamental step in the comprehension of the whole
phenomenology. As it can be easily imagined, the derivation of
caustic surfaces in a full general relativistic context is much
more involved than in classical lens models analyzed under the
weak deflection paradigm. The simplest general relativistic lens
is the Schwarzschild black hole. In this model, however, the
caustics are degenerate and are therefore trivially tractable. The
first general relativistic lens with a non-trivial caustic
structure is the Kerr black hole. Going beyond the results of
previous works, focused on particular limits of the caustics
\cite{RauBla,SerDeL,BDSS,BDS,BozSca}, this paper contains a
complete investigation of the full caustic structure of the Kerr
metric. This represents a considerable step forward for
gravitational lensing phenomenology in full General Relativity.

In summary, we have shown 3-dimensional pictures of the primary
and higher order caustic surfaces. We have analyzed the dependence
of the cross-sections of the caustic surfaces on the source
distance, the black hole spin, the spin inclination and the
caustic order. We have shown that the caustic surfaces always wind
an infinite number of times around the horizon following a
logarithmic law. The size of the caustics always remains finite in
the whole parameter space and there are no transitions to
different kinds of caustic singularities in the lens mapping. For
extremal spin values, the size of higher order caustics increases
exponentially with the caustic order. This implies that the number
of higher order images also grows exponentially, as the same
caustic can be crossed an exponential number of times. We have
compared our results with previous analytical and numerical
approximations, finding perfect agreement in the respective limits
of validity of the approximations. We have also shown some
critical curves in the observer's sky, focusing on their
dependence on the source distance.

In addition, the code developed in this paper for the calculation
of the caustics is particularly well-suited for the study of
higher order images, as being inspired by the Strong Deflection
Limit methodology. An interesting future development of this code
might be the implementation of an efficient resolution algorithm
of the Kerr lens equation including higher order images.

Besides the purely theoretical interest of the analysis presented
in this paper, we can easily imagine that the knowledge of the
complete caustic structure of the Kerr black hole in all ranges of
parameters will stand as an extremely helpful guide in future
astrophysical applications involving very strong light bending by
the gravitational field of spinning black holes.

\begin{acknowledgments}
The author thanks Gaetano Scarpetta and Mauro Sereno for useful
comments on the manuscript. We acknowledge support for this work
by MIUR through PRIN 2006 Protocol 2006023491\_003, by research
funds of Agenzia Spaziale Italiana, and by research funds of
Salerno University.
\end{acknowledgments}

\appendix
\section{From the geodesics equation to the lens mapping}

In this appendix, we describe all the steps of the calculation of
the caustics in Kerr spacetime. Section \ref{Basics} recalls the
basic facts about Kerr geodesics. Section \ref{Labels} discusses
the unstable photon orbits and the choice of good variables for
numerical integration. Section \ref{Angular} deals with angular
integrals, Section \ref{Radial} with radial integrals. Section
\ref{LensEq} introduces the $\psi$ variable and shows the lens
mapping explicitly. Finally, Section \ref{Numerical} discusses
details on the numerical implementation of the strategy for
searching critical points and drawing caustic surfaces.

\subsection{Basics of Kerr geodesics} \label{Basics}

Thanks to the separability of the Hamilton-Jacobi equation
\cite{Car}, the geodesic motion for a massless particle is
described by two integral equations

\begin{eqnarray}
&& \pm \int \frac{dr}{\sqrt{R}}=\pm \int \frac{d
\vartheta}{\sqrt{\Theta}} \label{Geod1}\\
& \phi_f-\phi_i =& a \int\frac{r^{2}+a^{2}-a J}{\Delta \sqrt{R}}
dx-a \int \frac{dx}{\sqrt{R}} \nonumber  \\
&& + J \int \frac{\csc^2\vartheta}{\sqrt{\Theta}} d \vartheta,
\label{Geod2}
\end{eqnarray}
where
\begin{eqnarray}
&\Theta=&Q+a^2 \cos^2\vartheta-J^2 \cot^2\vartheta \\
&R=&r^4+(a^2-J^2-Q)r^2+2Mr(Q+(J-a)^2) \nonumber \\ &&-a^2 Q,
\label{R}.
\end{eqnarray}

$J$ and $Q$ are two constants of motion. The first can be
identified with the projection of the angular momentum of the
particle on the spin axis and the second is related to the square
of the total angular momentum. The double signs before the
integrals are chosen in such a way that all pieces of the
integrals between two consecutive inversion points are summed with
a positive sign.

A photon reaching the observer with constants of motion $J$ and
$Q$ will then be detected by the observer at angular coordinates
\cite{Cha}
\begin{eqnarray}
&& \theta_{1}=- \frac{J}{r_o \sqrt{1-\mu_o^2}}, \label{Th1J} \\
&& \theta_{2}=\pm r_o^{-1}\sqrt{Q + \mu_o^2\left( a^2-
\frac{J^2}{1-\mu_o^2} \right) }.\label{Th2Q}
\end{eqnarray}

Only an indetermination on the sign of $\theta_2$ remains.

Conversely, the constants of motion of a photon reaching the
observer from angles $(\theta_1, \theta_2)$ are found by inverting
Eqs. (\ref{Th1J})-(\ref{Th2Q})
\begin{eqnarray}
&& J=-\theta_{1} r_o \sqrt{1-\mu_o^2}, \label{JTh1} \\
&& Q=\theta_{2}^2 r_o^2 +\mu_o^2(\theta_{1}^2
r_o^2-a^2).\label{QTh2}
\end{eqnarray}

\subsection{Choosing good labels for photon trajectories} \label{Labels}

All photon trajectories ending at the observer can be completely
specified by assigning proper values to the constants of motion
$J$ and $Q$. Equivalently, one can specify the coordinates in the
observer sky $\theta_1$ and $\theta_2$. However, in order to keep
numerical errors low, we may need to find new labels for the
photon trajectories. In the case of the Kerr metric, the most
challenging geodesics are those approaching the unstable photon
orbit. We have found that using variables inspired by the Strong
Deflection Limit (SDL) technique \cite{BDS} proves very
convenient, opening the way to a reliable description of higher
order caustics. Therefore, in this subsection we recall the
definition of the SDL parameters $\delta$ and $\epsilon$, and
introduce the variable $\eta$ which replaces the $\xi$ of Ref.
\cite{BDS}.

The inversion points in the radial motion are obtained by solving
Eq. $R=0$. Photons reaching an observer at infinity can have one
or zero radial inversion points. If $\partial R/\partial r =0$ is
satisfied at the same time with $R=0$ for some value $r=r_m$, then
the photon remains at radial coordinate $r_m$ forever. However,
small perturbations would make the photon fall into the black hole
or escape to infinity, because the circular photon orbits around
Kerr black holes are unstable. The unstable photon orbits are
characterized by particular values of the constants of motion $J$
and $Q$. In order to find them, we can solve equations $R=0$ and
$\partial R/\partial r =0$ for $J$ and $Q$ as functions of
$r=r_m$. As $r_m$ varies, $J$ and $Q$ describe a locus in the
$(J,Q)$ space, corresponding to all possible unstable photon
orbits. Explicitly, this locus is given by \cite{Cha}
\begin{eqnarray}
&&J_{m}(r_m)=\frac{(3M-r_m)r_m^2-a^2(r_m+M)}{a(r_m-M)} \label{Jm}\\
&&
Q_{m}(r_m)=\frac{r_m^3\left[4a^2M-r_m(r_m-3M)^2\right]}{a^2(r_m-M)^2}.
\label{Qm}
\end{eqnarray}

A photon orbiting at $r_m$ is thus characterized by constants
$J_m(r_m)$ and $Q_m(r_m)$. The allowed values of $r_m$ are those
keeping $Q$ positive. By perturbing the unstable orbit, we can
make the photon escape and reach a distant observer. In this case,
plugging Eqs. (\ref{Jm})-(\ref{Qm}) into Eqs.
(\ref{Th1J})-(\ref{Th2Q}), one finds that the escaped photon is
detected at position
\begin{eqnarray}
&\theta_{1,m}(r_m)&= \frac{r_m^2(r_m-3M)+a^2(r_m+M)}{r_o a(r_m-M)\sqrt{1-\mu_o^2}} \label{Shadow1}\\
&\theta_{2,m}(r_m)&=\pm \frac{\sqrt{\Lambda(r_m)}}{M r_o
a (r_m-M)\sqrt{1-\mu_o^2}} \label{Shadow2} \\
& \Lambda(r_m)&= r_m^3 \left[4M a^2-r_m(r_m-3M)^2 \right]
\nonumber \\
&& -2a^2r_m(2a^2M-3Mì2r_m+r_m^3)\mu_o^2 \nonumber
\\ && -a^4(r_m-M)^2\mu_o^4. \label{Lambda}
\end{eqnarray}

Not all perturbed photons can reach an observer out of the
equatorial plane. Only those photons with $r_m$ such that
$\Lambda(r_m)>0$ can be detected. Solving Eq. $\Lambda(r_m)=0$
numerically, we can find the two extrema of the allowed interval
for $r_m$ for any given values of the parameters $a$ and $\mu_o$.
We shall indicate these two extrema by $r_+$ and $r_-$.

Letting $r_m$ vary between $r_+$ and $r_-$ in Eqs. (\ref{Shadow1})
and (\ref{Shadow2}) and allowing both signs for $\theta_2$, we
obtain a closed curve in the observer's sky, which is usually
referred to as the border of the ``shadow'' of the black hole,
because the radiation deflected from the black hole would appear
outside this closed curve. Some shadows borders for different
valus of $a$ are shown in Fig. \ref{Fig shadow}. When the observer
lies on the equatorial plane ($\mu_o=0$), $r_+$ coincides with the
radius of the unstable photon orbit for equatorial prograde
photons, whereas $r_-$ becomes the radius of the unstable photon
orbit for equatorial retrograde photons, which is typically
larger. Increasing the inclination, $r_+$ and $r_-$ approach each
other until they coincide when the observer lies at the pole of
the black hole ($\mu_o=1$).

As $r_+$ and $r_-$ depend on the spin and the inclination, it is
convenient to introduce the variable $\eta$, related to $r_m$ by
\begin{equation}
r_m=\frac{1}{2}\left[r_+(1-\cos \eta)+r_-(1+\cos \eta) \right].
\label{rmeta}
\end{equation}
As $\eta$ varies from $0$ to $\pi$, $r_m$ varies from $r_-$ to
$r_+$. If we replace the double sign in Eq. (\ref{Shadow2}) by
$\mathrm{sign}[\eta]$, we can get the whole shadow border at once,
by varying $\eta$ in the range $[-\pi,\pi]$.

As anticipated in Section II, $\eta$ works as an angular
coordinate: $\eta=0$ corresponds to retrograde photons, appearing
on the right of the black hole; $\eta=\pi/2$ corresponds to
photons on nearly polar orbits, reaching the observer from above
the black hole; $\eta=\pi$ corresponds to prograde photons,
detected on the left of the black hole.

In addition, we introduce the new variable $\epsilon$ by the
relations
\begin{eqnarray}
&& \theta_1=\theta_{1,m}(\eta)(1+\epsilon) \label{th1par}\\
&& \theta_2=\theta_{2,m}(\eta)(1+\epsilon) \label{th2par}.
\end{eqnarray}
As $\epsilon$ varies in the range $[-1,+\infty)$ and $\eta$ in
$[-\pi,\pi]$, $\theta_1$ and $\theta_2$ span the whole observer
sky. This construction is very close to that presented in Ref.
\cite{BDS}, though here we make no expansion in powers of $a$. The
variables $\eta$ and $\epsilon$ can be used to label geodesics
ending at the observer and are particularly well-suited for
numerical calculations. The corresponding values of the constants
of motion $J$ and $Q$ can be found by plugging Eqs. (\ref{th1par})
and (\ref{th2par}) into Eqs. (\ref{JTh1}) and (\ref{QTh2}).

When $\epsilon>0$, Eq. $R=0$ admits a non-degenerate inversion
point $r_*$, whose expression can be written as
\begin{equation}
r_*=r_m(1+\delta).
\end{equation}

$\delta$ is a function of $\epsilon$ and $\eta$. It can be found
by solving Eq. $R=0$ numerically for given values of $\epsilon$,
$\eta$ and the parameters $a$ and $\mu_o$. For small values of
$\epsilon$, we have $\delta \sim \sqrt{\epsilon}$. However, for
low values of the inclination $\mu_o$ and for high values of the
spin, prograde photons ($\eta \simeq \pi$) tend to satisfy a
linear relation of the type $\delta \sim \epsilon$. However,
$\delta$ is not an observable and these relations are specific of
Boyer-Lindquist coordinates.

\subsection{Angular integrals} \label{Angular}

Eqs. (\ref{Geod1}) and (\ref{Geod2}) contain integrals on the
polar angle $\vartheta$. Performing the transformation $\cos
\vartheta =\mu$, they become
\begin{eqnarray}
&& J_1=\pm \int \frac{1}{\sqrt{\Theta_\mu}} d \mu \label{J1mu}\\
&& J_2=\pm \int \frac{1}{(1-\mu^2)\sqrt{\Theta_\mu}} d \mu
,\label{J2mu}
\end{eqnarray}
where
\begin{eqnarray}
& \Theta_\mu=&a^2(\mu_-^2+\mu^2)(\mu_+^2-\mu^2) \\
& \mu_\pm^2=&\frac{\sqrt{b_{JQ}^2+4a^2Q}\pm b_{JQ}}{2 a^2} \label{mupm}\\
& b_{JQ}=& a^2-J^2-Q.
\end{eqnarray}

The integration extrema are the source and observer polar
coordinates $\mu_s$ and $\mu_o$. However, for each inversion in
the polar motion, we must split the integration domain and change
sign.

The final result can be expressed in terms of elliptic integrals
of the first and third kind
\begin{eqnarray}
& J_1&(\mu_s,\epsilon,\eta)=\frac{1}{a\mu_-}\left[ -\mathrm{sign}[\eta]F(\lambda_o,k)+ \right. \nonumber \\
&& \left. (-1)^m \mathrm{sign}[\eta]F(\lambda_s,k) +2m K(k) \right] \\
& J_2&(\mu_s,\epsilon,\eta)=\frac{1}{a\mu_-}\left[ -\mathrm{sign}[\eta]\Pi(\mu_+^2,\lambda_o,k)+ \right. \nonumber \\
&& \left.(-1)^m \mathrm{sign}[\eta]\Pi(\mu_+^2,\lambda_s,k) +2m \Pi(\mu_+^2,k) \right]\\
& \lambda_s&=\arcsin \frac{\mu_s}{\mu_+} \\
& \lambda_o&=\arcsin \frac{\mu_o}{\mu_+} \\
& k&=-\frac{\mu_+^2}{\mu_-^2},
\end{eqnarray}
where $m$ is the number of inversions in the polar motion, and the
$\mathrm{sign}[\eta]$ takes into account the fact that for
positive $\eta$ the observer is reached from above and for
negative $\eta$ is reached from below. We recall the definitions
of the elliptic integrals
\begin{eqnarray}
&& F(\lambda,k)=\int\limits_0^\lambda \frac{d\vartheta}{\sqrt{1-k
\sin^2\vartheta
}}  \\
&& K(k)=F(\pi/2,k) \\
&& \Pi(n,\lambda,k)=\int\limits_0^\lambda
\frac{d\vartheta}{\left(1-n \sin^2\vartheta \right)\sqrt{1-k
\sin^2\vartheta
}}  \\
&& \Pi(n,k)=\Pi(n,\pi/2,k).
\end{eqnarray}

Besides depending on the source polar coordinate $\mu_s$, the
functions $J_1$ and $J_2$ also depend on $\epsilon$ (weakly) and
$\eta$ through the constants of motion $J$ and $Q$, appearing in
the expressions of $\mu_+$ and $\mu_-$.

\subsection{Radial integrals} \label{Radial}

The geodesics equations (\ref{Geod1}) and (\ref{Geod2}) also
contain two integrals on the radial coordinate
\begin{eqnarray}
&&I_1=\int \frac{dx}{\sqrt{R}} \\
&&I_2=\int \frac{dx}{\sqrt{R}} \int\frac{r^{2}+a^{2}-a J}{\Delta
\sqrt{R}} dx. \label{I2}
\end{eqnarray}

These integrals depend on the labels $(\epsilon,\eta)$ identifying
the geodesic through $J$ and $Q$.

The photon might be emitted from the source either outward or
inward. We must therefore distinguish these two cases and solve
the integrals accordingly.

If the photon is emitted in the outward direction, there is no
inversion point and the integrations are carried out in the domain
$[r_s,\infty)$. For numerical reasons, it is convenient to change
the integration variable to $z$, defined by
\begin{equation}
r=\frac{r_m}{1-z}.
\end{equation}
Then the integration range becomes $[1-r_m/r_s,1]$. In this case,
the integrand is sharply peaked at $z=0$ \cite{BozSca}.

If the photon is emitted inward, it must have an inversion point
$r_*$ somewhere between $r_m$ and $r_s$ in order to come back and
reach a distant observer. Therefore, this case is relevant for
lensing only if $\epsilon>0$ and $r_s>r_m$. The radial integrals
are split in two pieces: the approach piece, with extrema
$[r_*,r_s]$, and the escape piece, with extrema $[r_*,\infty)$.
The two pieces are conveniently calculated using the variable
\begin{equation}
r=\frac{r_*}{1-z},
\end{equation}
which maps the integration domains to the finite intervals
$[0,1-r_*/r_s]$ and $[0,1]$ respectively. In this case the
integrand diverges at $z=0$, but the result of the integration is
finite for any value of $\epsilon>0$. As $\epsilon \rightarrow 0$,
the integrals diverge logarithmically, except for prograde orbits
at high values of the spin, for which the integrals diverge as
$\epsilon^{-1}$.

For a given source at $r_s>r_m$, there exists a maximum value for
$\epsilon$ (we call it $\epsilon_{max}$) such that $R$ is positive
only for $\epsilon<\epsilon_{max}$. When
$\epsilon=\epsilon_{max}$, $r_s$ becomes a solution of the
equation $R=0$ and thus coincides with the inversion point $r_*$.

Summing up, the functions $I_1(\epsilon,\eta)$ and
$I_2(\epsilon,\eta)$ have two branches if $r_s>r_m$: one
corresponding to photons emitted outward, which exists for
$\epsilon$ in $[-1,\epsilon_{max}]$, and the other corresponding
to photons emitted inward, existing for $\epsilon$ in
$(0,\epsilon_{max}]$. The two branches nicely join at
$\epsilon_{max}$. If $r_s<r_m$, only the outward branch exists for
$\epsilon$ in $[-1,0)$.

\subsection{Lens mapping} \label{LensEq}

At this point, we introduce a new variable $\psi$, which
considerably simplifies numerical calculations and assumes a key
role in their success. This variable is defined by the relation
\begin{equation}
2K(k)\psi=a\mu_- I_1(\epsilon,\eta)
+\mathrm{sign}[\eta]F(\lambda_o,k). \label{psidef}
\end{equation}

The variable $\psi$ is directly defined in terms of the radial
integral $I_1$, thus eliminating the problems of the two branches
that this function has when is expressed in terms of $\epsilon$.
The reason behind the coefficient $2K(k)$ will be clear later.

\begin{figure}
\resizebox{\hsize}{!}{\includegraphics{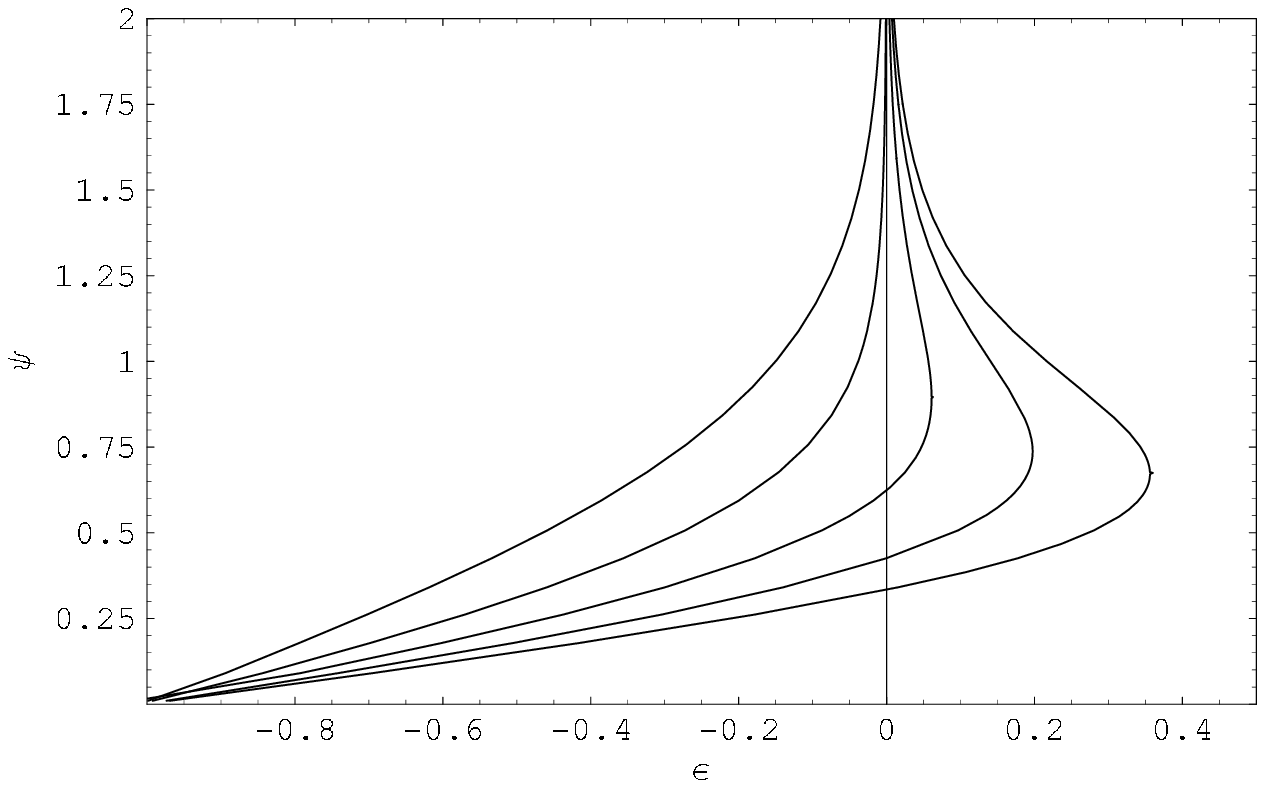}}
\caption{Relation between $\psi$ and $\epsilon$ for $a=0.2M$,
$\mu_o=0$, $\eta=\pi/4$. From left to right, the curves are drawn
for $r_s=2M,3M,4M,5M,6M$ respectively. Note that for these values
we have $r_m=3.16M$. }
 \label{Fig psieps}
\end{figure}

In Fig. \ref{Fig psieps} we plot $\psi$ vs $\epsilon$ for a
generic choice of the parameters. We note that when $r_s<r_m$,
$\epsilon<0$ for all values of $\psi$, whereas for $r_s>r_m$
$\epsilon$ reaches the maximum value $\epsilon_{max}$,
corresponding to $r_s=r_*$ and then tends to zero for very large
values of $\psi$. On the other hand, $\epsilon$ is a single-valued
function of $\psi$. Therefore, we can replace all occurrences of
$\epsilon$ by $\epsilon(\psi)$ obtained by solving Eq.
(\ref{psidef}) numerically.

Multiplying Eq. (\ref{Geod1}) by $a\mu_-$, separating the $\mu_s$
term and applying the Jacobi elliptic function $\mathrm{sn}$, we
get
\begin{equation}
\mu_s=\mathrm{sign}[\eta] \mu_+ \mathrm{sn}\left[2K(k)\psi,k
\right], \label{Lens1}
\end{equation}
which represents the lens equation for the polar coordinate
$\mu_s$.

The Jacobi elliptic function $\mathrm{sn}(x,k)$ is periodic in $x$
with period $4K(k)$. It tends to the usual trigonometric $\sin$
when $k$ is small (which is more or less our case, since
$\mu_+<\mu-$). So, the right hand side of Eq. (\ref{Lens1})
vanishes at $\psi=m$ with $m$ integer, is maximum at $\psi=2m+
1/2$, and minimum at $\psi=2m+ 3/2$. The interval in which there
are $m$ inversion points in the polar motion is $m- 1/2<\psi<m+
1/2$. As we shall see, the fact that this interval has a trivial
form in terms of $\psi$ for any values of $m$ is crucial in the
search for the critical points of higher order. If we had kept
$\epsilon$, the definition of the search domain would have been
much more problematic. In Ref. \cite{BDS}, the variable $\psi$
played a similar role, but its definition was restricted to the
limit of small black hole spin. In that case $\psi$ reduces to the
equivalent deflection by a Schwarzschild black hole of the same
mass. Eq. (\ref{psidef}) generalizes the former definition and
allows a fruitful use of this variable in all the Kerr parameter
space.

As regards the azimuthal motion, the extrema of $\phi$ are
$\phi_i=\phi_s$ and $\phi_f=\phi_o$. Recalling that we have set
$\phi_o=0$, Eq. (\ref{Geod2}) can be written in the form
\begin{equation}
\phi_s= a\left[ I_1(\psi,\eta)-I_2(\psi,\eta) \right]-J
J_2(\mu_s,\psi,\eta). \label{Lens2}
\end{equation}
This is the lens equation for the azimuthal coordinate $\phi_s$.
Along with (\ref{Lens1}), it constitutes the lens mapping for a
Kerr black hole in the form of Eqs. (\ref{SimLens1}) and
(\ref{SimLens2}).

\subsection{Remarks on numerical implementation} \label{Numerical}

Here we give a few technical details regarding the numerical
implementation of the algorithm for finding critical curves and
caustics.

As input parameters we consider the black hole spin $a$, the
inclination $\mu_o$, the source distance $r_s$, the caustic order
$m$ and the value of the variable $\eta$. Of course, depending on
which kind of plot we want to draw, we may decide to cycle on
different variables. For example, by cycling on $\eta$ keeping all
other parameters fixed, we get a cross-section of the caustic
surface at fixed source distance. Cycling on $a$, while keeping
$\eta$ at some specific value, we may get the position of a cusp
(for example) as a function of the black hole spin.

Starting from each set of parameters $\{a,\mu_o,r_s,m,\eta\}$,
first we calculate $r_+$ and $r_-$ by solving Eq. $\Lambda(r)=0$
with $\Lambda(r)$ given by Eq. (\ref{Lambda}). Then we find $r_m$
by Eq. (\ref{rmeta}).

For any test value of $\psi$, we numerically invert Eq.
(\ref{psidef}) to find $\epsilon$, keeping track of the branch of
$I_1$ in which we are. If we are in the branch characterized by
the presence of the radial inversion point, we also need to
calculate $\delta$ by numerical solution of Eq. $R=0$.

All these equations are solved using the Mathematica FindRoot
routine with the secant method. The integrations of $I_1$ and
$I_2$ are performed using the NIntegrate routine, which works
pretty well on the finite domains obtained by changing to the $z$
integration variable.

At this point, we can evaluate the lens equations (\ref{Lens1})
and (\ref{Lens2}) to get $\mu_s$ and $\phi_s$, without further
effort.

The Jacobian of the lens mapping is obtained by evaluating
difference quotients of $\mu_s$ and $\phi_s$ with respect to
$\psi$ and $\eta$.

Finally, the zero of the Jacobian at fixed $\eta$ is found by
applying the secant method with respect to $\psi$ in the range
$[m-1/2,m+1/2]$.

Depending on the required precision and the number of points in
the cycle, the evaluation of a caustic may take several minutes.
We have required a precision of $10^{-6}$ in $\psi$ and about 500
points for each caustic, refining the sampling in the neighborhood
of the cusps. With this setup, a cross-section at fixed $r_s$
takes about 5 minutes on a laptop. Numerical noise has been
detected on prograde orbits at extremal spin only starting from
the fifth order caustic.

\end{document}